\documentclass[11pt,letterpaper]{article}
\usepackage[T1]{fontenc}
\usepackage{lmodern}
\usepackage{amsmath,amssymb,amsthm}
\usepackage[margin=2.5cm]{geometry}
\usepackage{graphicx,booktabs,tabularx,array,float,enumitem,xcolor}
\usepackage[hidelinks]{hyperref}
\hypersetup{pdftitle={Global saddle selection and exact absolute-instability boundaries for convection in porous media with throughflow},
            pdfauthor={Oleg N. Kirillov}}
\usepackage[numbers,sort&compress]{natbib}
\usepackage{placeins}
\newcommand{\ii}{\mathrm{i}}
\newcommand{\Ra}{\mathit{Ra}}
\newcommand{\Pe}{\mathit{Pe}}
\newcommand{\Rey}{\operatorname{Re}}
\newcommand{\Imy}{\operatorname{Im}}
\newtheorem{theorem}{Theorem}
\newtheorem{lemma}[theorem]{Lemma}
\newtheorem{proposition}[theorem]{Proposition}

\newcolumntype{L}{>{\raggedright\arraybackslash}X}
\title{Global saddle selection and exact absolute-instability boundaries\\ for convection in porous media with throughflow}
\author{Oleg N.~Kirillov\\[3pt]
\small Northumbria University, Newcastle upon Tyne, UK\\
\small\href{mailto:oleg.kirillov@northumbria.ac.uk}{oleg.kirillov@northumbria.ac.uk}}
\date{}
\begin{document}
\maketitle
\begin{abstract}
In the Darcy--Prats problem, horizontal throughflow leaves the temporal onset of convection unchanged but raises the
heating required for a localized disturbance to grow at a fixed point. The saddle equations give five candidates,
but only the saddles reached by deformation of the Fourier contour contribute to the growth. For $\Ra>\pi^2$, $\Pe>0$
we prove that the same off-axis pair contributes throughout the domain: no Stokes transition can change its contribution.
Its growth gives a unique analytic absolute threshold,
first reached by the lowest vertical mode. Rescaling transfers the selection to Forchheimer drag, transverse rolls of a
power-law fluid, anisotropic permeability, an open upper boundary and a vertical magnetic field. A shift of the same
growth function accounts for interphase heat exchange in a model of metallic foam. At prescribed
throughflow the threshold increases strictly with either resistance coefficient. In particular, a vertical magnetic
field raises the absolute threshold at fixed P\'eclet number. Tabulated thresholds check the reduction.
Algebraic elimination gives the boundary polynomial, and frequency recovery selects its physical branch. At the
Darcy level the selected saddles trace an arc of the elliptic curve 11a3, a degeneration of a genus-two family at
nonzero shifted levels. Series inversion on this arc gives an integer weak-throughflow series and an all-order strong-throughflow
expansion. Rational certificates establish the exact convergence radii of both series on the selected sheets.
\end{abstract}


\section{Introduction}\label{sec:intro}

In a porous layer heated from below, buoyancy amplifies a temperature disturbance and diffusion smooths it. A horizontal throughflow, measured by the P\'eclet number $\Pe$, adds advection. A disturbance may grow while it travels and still leave the region where it was created, so a growing plane
wave does not mean that a fixed probe records a growing signal. Prats~\cite{Prats1966}
showed that a uniform throughflow leaves the temporal onset at the Darcy--Rayleigh number
$\Ra=4\pi^2$ unchanged. Spatio-temporal analysis reveals a second threshold, the transition from convective to absolute instability~\cite{Briggs1964,Bers1983,HuerreMonkewitz1990,Chomaz2005}.

For convection in a porous layer with throughflow, Dufour and N\'eel~\cite{DufourNeel1998} used a near-onset
amplitude equation to study traveling rolls and fronts and to obtain the leading correction to the absolute threshold.
Chung, Kim and Choi~\cite{ChungKimChoi2009} derived the same weak-throughflow correction.
Delache, Ouarzazi and Combarnous~\cite{Delache2007,Delache2005} computed absolute thresholds by continuing
spatial branches and compared them with experiment; see also~\cite{DiazBrevdo2011,HirataOuarzazi2010}.
Barletta and Alves~\cite{BarlettaAlves2017}, Barletta and Celli~\cite{BarlettaCelli2017} and the monograph of
Barletta~\cite{Barletta2019} made the selection mechanism explicit and turned it into a systematic method for porous-medium convection. Some solutions of the neutral-saddle conditions must be
discarded, because the corresponding contour cannot be obtained from the real Fourier contour. These works give the neutral curves, the limiting complex wavenumber, the leading coefficient $c_1$ at strong throughflow, and the exclusion of a spurious stationary parabola through the holomorphy obstruction of a pole, also
at positive $\Pe$. Barletta and Celli write the curve of the off-axis neutral saddles in closed polar form.
Turkyilmazoglu~\cite{Turkyilmazoglu2023} parametrizes the transverse neutral saddle locus in radicals for a vertical
magnetic field, including the nonmagnetic Prats limit, and gives the frequency once the saddle is chosen.
This program has also been extended by Barletta, Celli, Alves and coworkers to a power-law fluid~\cite{AlvesBarletta2015}, anisotropic permeability~\cite{CelliBarletta2022}, anomalous
diffusion~\cite{Barletta2026}, rectangular channels~\cite{Barletta2020channel} and symmetric wall heat fluxes with an inclined throughflow~\cite{Celli2025}.
Freitas \emph{et al.}~\cite{Freitas2022} computed absolute thresholds for a porous matrix that exchanges heat with the fluid.
These studies supply independent benchmarks.

Previous studies obtained candidate thresholds by numerical saddle continuation or by explicit neutral-saddle formulae.
In the numerical route one solves the three real equations $\Rey\lambda=0$, $\partial\lambda/\partial k=0$ for the
unknowns $(\Rey k,\Imy k,\Ra)$, where $\lambda(k)$ is the complex growth rate of a mode $e^{\lambda t+\ii kx}$. Several
roots appear, and the smallest admissible one above the temporal onset is kept. Admissibility requires a steepest-descent
deformation of the Fourier contour through the saddle without crossing a singularity~\cite[Sec.~3.5.3]{Barletta2019}.
Suslov~\cite{Suslov2006} discusses numerical continuation. Alves \emph{et al.}~\cite{AlvesHirataSchuabbBarletta2019}
use adjoint sensitivities to locate candidate saddles and facilitate the causality check in differential eigenvalue problems.
For the Prats problem, Barletta~\cite[Sec.~8.2.4, p.~190]{Barletta2019} traces the neutral pair from $\Pe=0$ and notes that
in practice the contour is checked at finitely many values of $\Pe$; see also~\cite{Freitas2022}.
Continuity of a saddle does not guarantee that it continues to contribute. We prove that the same pair contributes throughout $\Ra>\pi^2$, $\Pe>0$, including away from the neutral curve.
The largest growth rate is taken over contributing saddles~\cite[App.~B]{Deissler1987}.

Admissibility is a topological property. A saddle contributes to the large-time asymptotics of the Fourier integral when its steepest-ascent path, the cycle dual to its Lefschetz thimble, has a nonzero intersection number with the original
contour~\cite{Pham1967,Pham1983,BerryHowls1991,DelabaereHowls2002}. For the simple saddles and the physical contour
considered here, pinching in the sense of Briggs and Bers is the local form of this statement. In general, pointwise growth and algebraic pinched-double-root criteria must be distinguished~\cite{HolzerScheel2014}.
Intersection numbers change only at Stokes phenomena or saddle coalescences~\cite{HallHeckrotte1968,KupferBersRam1987}. For the Prats family and its scaled generalizations, we prove that the contributing indices remain unchanged on the whole domain. The argument makes three selections: of the saddles by
the Fourier contour (section~\ref{sec:selection}), of the physical boundary by frequency recovery
(section~\ref{sec:algebraic}) and of the analytic sheets for the convergence radii (theorem~\ref{thm:radii}).
Table~\ref{tab:workflow} compares the two routes.

\begin{table}[!htbp]
\centering\footnotesize
\begin{tabularx}{\textwidth}{@{}lLL@{}}
\toprule
Step & Saddle continuation with admissibility checks & This work\\
\midrule
Saddles & Newton solution of $\Rey\lambda=\partial\lambda/\partial k=0$ at each $(\Pe,\dots)$ & all five saddles classified once for the whole domain (section~\ref{sec:saddles}) \\
Admissibility & holomorphy requirement checked on contour plots at selected $\Pe$; criteria $\Ra>0$, $\Ra>\Ra_c$ &
intersection indices proved constant on the whole domain (lemma~\ref{lem:invariance}, theorem~\ref{thm:selection})\\
Threshold & root of the selected branch, computed separately for each model & unique root of one monotone growth function for the whole class (theorem~\ref{thm:onset}, proposition~\ref{prop:level})\\
Closed forms & $c_1$, limiting wavenumber, polar and radical formulae for neutral saddles & exact boundary polynomial with
physical branch selection; elliptic model and all-order expansions; certified radii\\
\bottomrule
\end{tabularx}
\caption{The established route to absolute thresholds, used to obtain the published thresholds considered here, and what the
global selection adds.}
\label{tab:workflow}
\end{table}

Theorem~\ref{thm:selection} classifies all five complex saddles of the Prats problem for all $\Ra>\pi^2$, $\Pe>0$. The selected pair defines a growth function whose zero set is a
unique, analytic and strictly increasing threshold, with the lowest vertical mode minimizing it (theorem~\ref{thm:onset}).
Section~\ref{sec:heuristics} separates three sources of ambiguity: noncontributing saddles, roots introduced by frequency elimination and neutral imaginary saddles below the threshold. Algebraic elimination with frequency recovery gives the exact boundary and its crossing with the spurious parabola, $\Pe_\times=\frac\pi2\sqrt{71+17\sqrt{17}}$
(section~\ref{sec:algebraic}). A change of scale reduces Forchheimer drag, power-law transverse rolls, anisotropic
permeability, an open upper boundary, vertical magnetic resistance and interphase heat exchange to levels $s$ of one growth function. Each threshold
is a single monotone root (section~\ref{sec:universal}, algorithm~\ref{alg:onset}), and increases with either resistance
coefficient at prescribed throughflow (section~\ref{sec:magnetic}). For $s=0$ the selected neutral saddles lie on the elliptic curve 11a3, a degeneration of a genus-two family for
$s\ne0$. On this smooth model the selected arc gives all coefficients of the large-$\Pe$ series, an
integer small-$\Pe$ series and certified convergence radii (section~\ref{sec:geometry}).

\section{Wavepacket, throughflow and two thresholds}\label{sec:model}

\begin{figure}[!htbp]
\centering\includegraphics[width=\textwidth]{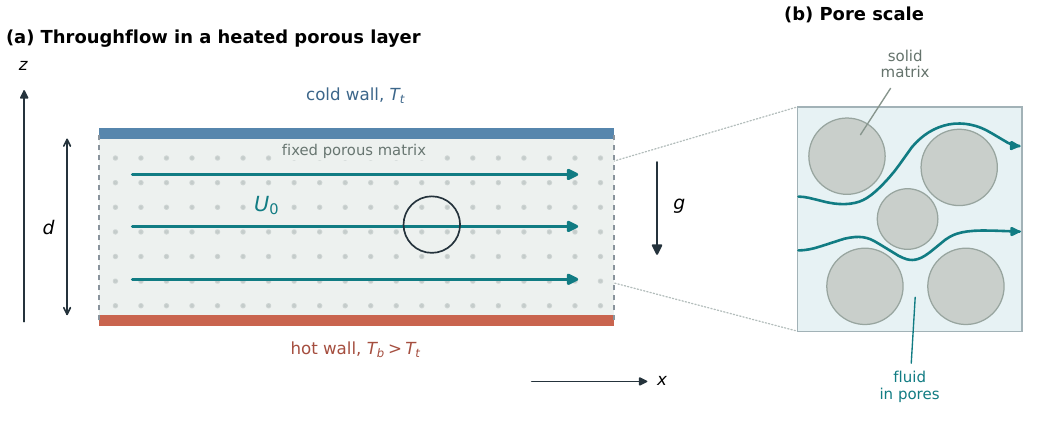}
\caption{Physical model. (a) A section of a horizontally unbounded porous layer of thickness $d$. The fluid seeps to the
right with mean Darcy velocity $U_0$; the lower impermeable wall is hotter than the upper one, and gravity points downwards.
(b) Close-up: the fluid flows through connected pores between fixed grains. Streamlines are schematic; the Darcy model
averages the pore structure. The equations use the dimensionless height $Z=z/d$. The figure was drawn with the AI-assisted Python code described under \hyperref[sec:computational-methods]{Computational methods}.}
\label{fig:model}
\end{figure}

The layer is unbounded in the horizontal coordinate $x$ and bounded by walls at $Z=0,1$. The walls are impermeable and
isothermal, and the lower wall is hotter. The basic state is a uniform horizontal throughflow with a linear temperature profile.
The temperature perturbation $\theta$ is measured from this profile, and the stream function $\psi$ describes the velocity
perturbation. The fluid moves through the pores of a fixed solid matrix under an imposed horizontal pressure gradient
(figure~\ref{fig:model}). In the basic state the temperature varies only vertically, so the throughflow does not advect its
gradient.

With the thermal time scale the linear equations read
\begin{equation}
 \nabla^2\psi=-\Ra\,\partial_x\theta,\qquad
 \partial_t\theta+\Pe\,\partial_x\theta+\partial_x\psi=\nabla^2\theta,
 \postdisplaypenalty=10000
 \label{eq:pde}
\end{equation}
where $\Ra$ and $\Pe$ are the Darcy--Rayleigh and P\'eclet numbers, respectively.

A positive heat-capacity factor is absorbed in the time scale, and the conventions follow~\cite{BarlettaAlves2017,Barletta2019}. The first equation is the instantaneous Darcy
response of the velocity to buoyancy. In the second, the advection of the basic temperature gradient competes with diffusion
and horizontal drift. The mode $(\psi,\theta)\propto\sin(m\pi Z)\,e^{\lambda t+\ii kx}$, $m=1,2,\ldots$, with complex growth rate $\lambda$ gives the exact
dispersion relation
\begin{equation}
 \lambda_m(k)=\Ra\frac{k^2}{k^2+m^2\pi^2}-(k^2+m^2\pi^2)-\ii\Pe k .
 \label{eq:dispersion}
\end{equation}
The fraction, which comes from eliminating the stream function, describes buoyant amplification. The second term is
diffusive loss and the third is advection of the packet.

\subsection{Temporal threshold}

For a real wavenumber the advection term in~\eqref{eq:dispersion} is purely imaginary: $\Imy\lambda_1(k)=-\Pe\,k$
only shifts the frequency, and the temporal growth rate of the first mode is
$\sigma(k)=\Rey\lambda_1(k)=\Ra\,k^2/(k^2+\pi^2)-(k^2+\pi^2)$. Its maximum over real $k$ is
\begin{equation}
 \sigma_{\max}=\Ra-2\pi\sqrt{\Ra},\qquad k_{\max}^2=\pi\sqrt{\Ra}-\pi^2,\qquad \Ra>\pi^2,
\end{equation}
while for $\Ra\le\pi^2$ the maximum $-\pi^2$ is attained at $k=0$. Temporal instability therefore starts at $\Ra=4\pi^2$,
independently of $\Pe$: the throughflow changes the frequencies of real waves but not their growth rates.

\begin{figure}[!htbp]
\centering\includegraphics[width=\textwidth]{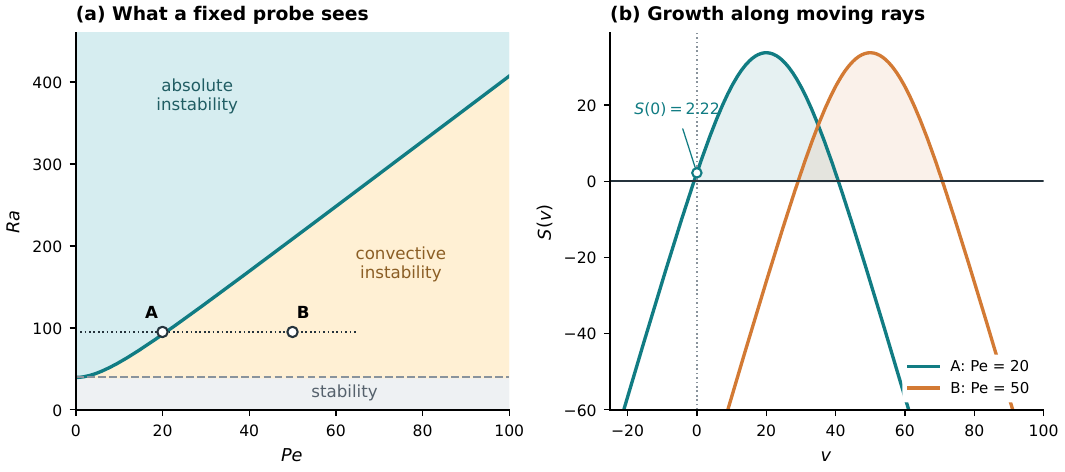}
\caption{Drift of a growing packet. (a) The temporal threshold $4\pi^2$ and the absolute threshold separate three regions.
Points A and B share $\Ra=95$ but have different throughflows. (b) At $\Pe=20$ the fixed point lies inside the growing
packet, $S(0)=2.2195$; at $\Pe=50$ the whole packet is swept downstream. The maximal temporal growth is the same. The figure was drawn with the AI-assisted Python code described under \hyperref[sec:computational-methods]{Computational methods}.}
\label{fig:packet}
\end{figure}

A localized disturbance requires the sum over all real wavenumbers. The temperature Green function of the first mode contains
\begin{equation}
 G(x,t)=\frac1{2\pi}\int_{\mathbb R}e^{\lambda_1(k)t+\ii kx}\,dk .
 \label{eq:green}
\end{equation}
On the ray $x=vt$ the integrand of~\eqref{eq:green} becomes $e^{t\lambda_v(k)}$, where
\[
 \lambda_v(k)=\lambda_1(k)+\ii vk=\Ra\frac{k^2}{k^2+\pi^2}-(k^2+\pi^2)-\ii(\Pe-v)k
\]
is the complex growth rate seen by an observer moving with speed $v$; $v=0$ corresponds to a fixed probe. Let
\[
 S(v)=\limsup_{t\to\infty}\frac1t\log|G(vt,t)|
\]
be the exponential rate along the ray. The $\limsup$ (equivalently, a local norm over a bounded window) accounts for the
oscillatory zeros produced by conjugate contributions~\cite{HolzerScheel2014}. If $S(0)>0$ the packet grows at a fixed point and the instability is absolute; if $S(0)<0$ but
$S(v)>0$ in some moving frame, it is convective. Since a change of observer only replaces $\Pe$ by $\Pe-v$, the spreading of the packet and the absolute threshold are
described by one function. Figure~\ref{fig:packet} shows both outcomes at the same heating. The spatial response of
the same flow to a time-periodic source is studied by Barletta~\cite{Barletta2021Space}.

For $\Ra>4\pi^2$, $S(v)$ is positive on an interval of observer speeds centered at $v=\Pe$ (figure~\ref{fig:packet}(b)).
The ends of this interval are the edges of the growing packet. The instability is absolute when the interval contains
$v=0$, so the trailing edge moves upstream. A stationary trailing edge marks the threshold. At $\Ra=95$ and $\Pe=20$ (point~A in
figure~\ref{fig:packet}(a)) the fixed point lies inside the growing packet, and $S(0)=2.2195$. At the same heating and
$\Pe=50$ (point~B) the whole packet is swept downstream. Both cases have the same maximal temporal growth rate.
Section~\ref{sec:edge} shows that the interval is $|\Pe-v|<c_*(\Ra)$ with a unique speed $c_*$.

\FloatBarrier
\subsection{Canonical variables}

For the first mode and an observer moving with speed $v$ we introduce
\begin{equation}
 \kappa=\frac{k}{\pi},\quad r=\frac{\Ra}{\pi^2},\quad p=\frac{|\Pe-v|}{\pi},\quad \Lambda=\frac{\lambda_v}{\pi^2},
 \label{eq:canonical}
\end{equation}
so that $\Lambda$ is the complex growth rate $\lambda_v$ in units of $\pi^2$. A reflection of the horizontal coordinate allows
$p>0$. The normalized symbol is
\begin{equation}
 L(\kappa;r,p)=r\frac{\kappa^2}{1+\kappa^2}-(1+\kappa^2)-\ii p\kappa .
 \label{eq:L}
\end{equation}
Multiplying $\Lambda-L=0$ by the denominator gives the spatial quartic
\begin{equation}
 D(\kappa,\Lambda)=(\Lambda+1+\kappa^2+\ii p\kappa)(1+\kappa^2)-r\kappa^2 .
 \label{eq:D}
\end{equation}
For a localized disturbance the relevant stationary points lie in the complex wavenumber plane, and one must decide which
complex saddles, the zeros of $L_\kappa=\partial L/\partial\kappa$, take part in the original
Fourier integral (a subscript $\kappa$, $r$ or $p$ on $L$, $D$, $F$ or $g$ denotes a partial derivative, as in
$L_\kappa$, $D_{\kappa\kappa}$, $F_r$ and $g_p$, and a prime denotes the derivative of a function of one variable, as
in $r_0'(p)$; all other subscripts are labels). Table~\ref{tab:notation} relates the canonical variables to the notation of~\cite{BarlettaAlves2017,Barletta2019}. Its lower block lists the notation introduced in section~\ref{sec:universal}
for the general symbol.

\begin{table}[!htbp]
\centering\small
\begin{tabularx}{\textwidth}{@{}lLl@{}}
\toprule
Symbol & Meaning & Canonical variable\\
\midrule
$\Ra$ ($R$ in~\cite{Barletta2019}) & Darcy--Rayleigh number & $r=\Ra/\pi^2$\\
$\Pe$; $v$ & P\'eclet number of the throughflow; speed of the observer & $p=|\Pe-v|/\pi$\\
$k$ & horizontal wavenumber & $\kappa=k/\pi$\\
$\lambda$; $\lambda_v=\lambda+\ii vk$ & complex growth rate in the laboratory frame and for an observer moving with speed $v$
($\Rey\lambda=\sigma$ for real $k$) & $\Lambda=\lambda_v/\pi^2$\\
\midrule
\multicolumn{3}{@{}l}{\emph{Introduced in section~\ref{sec:universal} for the general symbol~\eqref{eq:general}}}\\
$\mathcal R$, $\mathcal P$ & buoyancy and advection parameters & $r=\mathcal R/(A\pi^2)$, $p=\mathcal P/\alpha$\\
$A$, $B$ & coefficients of the buoyancy denominator $A\pi^2+Bk^2$ & $\alpha=\pi\sqrt{A/B}$, $\kappa=k/\alpha$\\
$s$ & level of the shifted problem, $s=B/A-1>-1$ & $s=0$ for Darcy\\
$g(r,p)$ & canonical growth rate of the selected saddle (theorem~\ref{thm:selection}) & threshold $g(r,p)=s$\\
\bottomrule
\end{tabularx}
\caption{Notation. Upper block: variables of~\cite{BarlettaAlves2017,Barletta2019} and the canonical variables~\eqref{eq:canonical}
of the Darcy--Prats problem. Lower block: notation of section~\ref{sec:universal}, which reduces to the upper block for $A=B=1$.}
\label{tab:notation}
\end{table}

\begin{figure}[!t]
\centering\includegraphics[width=\textwidth]{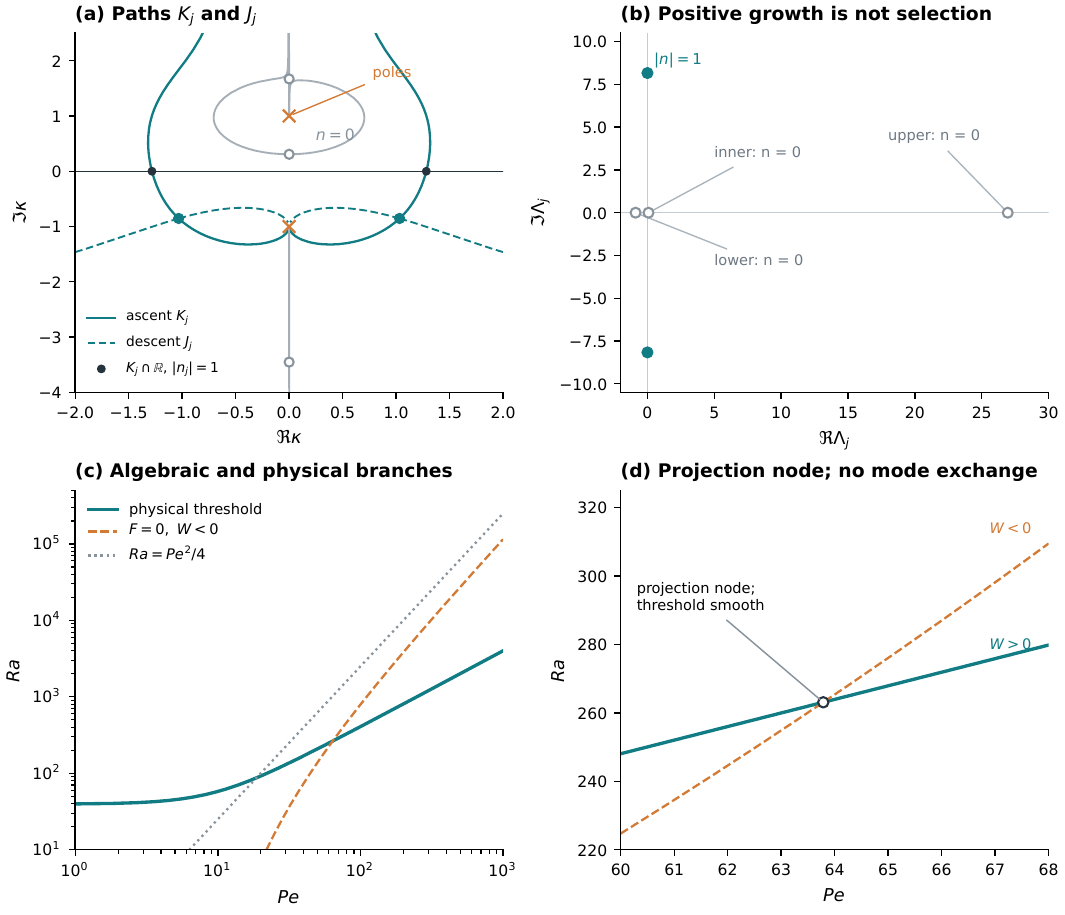}
\caption{Selection and its algebraic trace. (a,b) At $\Pe=20$, $\Ra=91.952768$: physical saddles filled, imaginary saddles
open, poles marked by crosses; panel (a) shows the steepest-ascent paths $K_j$ of all
five saddles (solid) and the steepest-descent paths $J_j$ of the off-axis pair (dashed); dots mark the single crossing of
each $K_j$ of the pair with the real contour. For the aligned imaginary saddles the ray is tilted sideways by $0.002$. (b) Critical values
$\Lambda_j=L(\kappa_j)$ of the five saddles: on the threshold the off-axis pair has $\Lambda_j=\mp8.165\,\ii$, a purely
oscillatory response, whereas the inner and upper imaginary saddles have positive real critical values, $0.079$ and
$26.945$, but zero index; the lower one has $-0.890$. A larger growth does not select a saddle. (c) Physical boundary, the extra real branch of
$F(r,P)=0$, and the spurious parabola. Here $P=p^2/4$, and $F$ is the boundary polynomial introduced in
section~\ref{sec:algebraic}, equation~\eqref{eq:resultant}. On the extra branch the squared frequency $W=\Omega^2$ of the neutral wave $\Lambda=\ii\Omega$ is negative, so it carries no
real oscillation. (d) Projection node: the physical threshold passes smoothly because the second
branch has $W<0$. The figure was drawn with the AI-assisted Python code described under \hyperref[sec:computational-methods]{Computational methods}.}
\label{fig:selection}
\end{figure}

\section{Five saddles and global selection}\label{sec:selection}

We prove that on the whole domain $r>1$, $p>0$ the physical growth is determined by the same pair of saddles. The domain
$r\le1$ needs no analysis: on the real contour $\Rey L\le-1$, so a localized disturbance cannot grow absolutely.

\subsection{Location of the saddles}\label{sec:saddles}

The saddles are the zeros of $L_\kappa=2r\kappa/(1+\kappa^2)^2-2\kappa-\ii p$. Multiplied by $-(1+\kappa^2)^2$,
which does not vanish away from the poles $\kappa=\pm\ii$, the stationarity condition $L_\kappa=0$ becomes the quintic
\begin{equation}
 (2\kappa+\ii p)(1+\kappa^2)^2-2r\kappa=0 .
 \label{eq:quintic}
\end{equation}
Its five roots are shown in figure~\ref{fig:selection}(a) for $\Pe=20$ on the threshold. Equivalently, $\kappa$ is a double root of the spatial quartic: along $\Lambda=L(\kappa)$ one has
$D_\kappa=-(1+\kappa^2)L_\kappa$, so $D=D_\kappa=0$ exactly at the zeros of~\eqref{eq:quintic}.

For a saddle on the imaginary axis, $\kappa=\ii y$, it becomes the real equation
\begin{equation}
 p=2y\left[\frac{r}{(1-y^2)^2}-1\right].
 \label{eq:imaginary}
\end{equation}
The $y$-derivative of the right-hand side, $2r(1+3y^2)(1-y^2)^{-3}-2$, is positive inside $(-1,1)$ and negative on each outer
interval when $r>1$. The limiting values show that there are exactly three simple imaginary saddles,
\begin{equation}
 \kappa_-=\ii y_-,\quad\kappa_0=\ii y_0,\quad\kappa_+=\ii y_+,\qquad y_-<-1<0<y_0<1<y_+ .
 \label{eq:three}
\end{equation}
The remaining two roots are symmetric with respect to the imaginary axis,
\begin{equation}
 \kappa_*=x_*-\ii\eta,\qquad-\overline{\kappa_*},\qquad x_*,\eta>0 .
 \label{eq:pair}
\end{equation}
As $p\downarrow0$ they emerge from the real maxima $\kappa=\pm\sqrt{\sqrt r-1}$. They cannot cross the real axis, where
$\Imy L_\kappa=-p\ne0$, and they cannot merge on the imaginary axis, because the three solutions of~\eqref{eq:imaginary} are
simple. All five saddles are therefore simple on the whole domain. Figure~\ref{fig:selection}(a) shows them at $\Pe=20$ on the threshold: the
pair~\eqref{eq:pair} as filled circles and the imaginary saddles~\eqref{eq:three} as open circles.

Each saddle is a square-root branch point of the spatial roots $\kappa(\Lambda)$, or a spatial exceptional point of order
two (EP2)~\cite{Kirillov2021}. A branch point contributes only when the Fourier contour admits the appropriate steepest-descent
deformation~\cite{ONaraighSpelt2013}.

\subsection{How a saddle enters the Green function}

A root of the quintic does not by itself determine the growth of the packet. From the critical value $\Lambda_j=L(\kappa_j)$, the value of $L$ at the saddle $\kappa_j$, we lift the ray $\Lambda=\Lambda_j+\rho$, $\rho\ge0$, to the surface $D=0$. Two spatial branches are born from the double
root, and their position relative to the original real contour decides whether the saddle enters the Green function.
Equivalently, let $J_j$ be the steepest-descent path through the saddle, its Lefschetz thimble, and $K_j$ the
steepest-ascent path, the dual cycle~\cite{Pham1967,Pham1983,BerryHowls1991}. The coefficient of the saddle in the decomposition of the real contour is the
oriented intersection number $n_j=\langle\mathbb R,K_j\rangle$: $n_j=0$ means no contribution and $|n_j|=1$ one contribution to the deformed contour. Figure~\ref{fig:selection}(a) shows these paths at $\Pe=20$: each
ascending path of the off-axis pair crosses the real axis once, while those of the imaginary saddles do not cross it.
The same contour coefficient defines an active saddle in the waveguide analysis of
Shanin, Korolkov and Kniazeva~\cite{ShaninKorolkovKniazeva}. Their test follows the two reverse steepest-descent
streams and accounts for cancellation when they hit the initial contour with opposite orientations. The zero level
used in their test relies on a real phase on that contour. Here the physical Fourier contour has nonconstant
$\Rey L$, and we use its oriented intersection with $K_j$ directly.

Lingwood~\cite{Lingwood1997} compares the Briggs test with steepest descent and gives examples of saddles that the
deformed Fourier contour cannot reach. In the construction of Briggs and Bers, the two spatial branches that merge in a contributing saddle lie on opposite sides
of the real $\kappa$-axis when $\Rey\Lambda$ is large, before the Laplace contour is
lowered~\cite{Briggs1964,Bers1983,HuerreMonkewitz1990}. If the temporal symbol $\lambda(k)$ is a polynomial in $k$, all spatial roots tend to infinity as $\Rey\Lambda\to+\infty$, so the two sides correspond to roots that go to infinity in opposite half-planes. The rational
symbol has poles at $\kappa=\pm\ii$: from $D=0$, two roots tend to them, $\kappa\approx\pm\ii\,(1+r/2\Lambda)$, and two tend to
infinity, $\kappa\approx\pm\ii\sqrt\Lambda$. Each side of the contour thus receives one root from a pole and one from infinity,
and the sides must be read off from the real contour rather than from the behavior at infinity.

Consider a continuous family of rational phases whose saddles remain simple. Assume that the pole orders and the decay
ends, the sectors near the poles and at infinity where $\Rey L\to-\infty$, persist, and that no saddle reaches a pole or
infinity. The integration contour is then transported in the corresponding relative homology. A chosen saddle has a
constant intersection index if its critical value never shares its imaginary part with another critical
value~\cite{BerryHowls1991,DelabaereHowls2002}. Indeed, $\Imy L$ is constant along gradient trajectories, so no
Stokes connection can involve that saddle. If a conjugate pair has nonreal critical values and all remaining critical
values are real, both members satisfy this condition.

\begin{lemma}[Invariance of the physical indices]\label{lem:invariance}
On the domain $r>1$, $p>0$ the critical values of the off-axis pair have nonzero imaginary parts of opposite sign. They
therefore never align with each other or with the critical value of an imaginary saddle, which is real, and the intersection
indices of the off-axis pair are constant on the domain. The indices of the imaginary saddles are determined directly in the
proof of theorem~\ref{thm:selection}.
\end{lemma}

\begin{proof}
Suppose the critical value of the pair were real. The symmetry $L(-\overline\kappa)=\overline{L(\kappa)}$ would then give two
distinct double roots of the quartic at the same real $\Lambda$. With $q=|\kappa_*|^2$ and $y=\Imy\kappa_*$ the quartic would
be the perfect square $D(\kappa,\Lambda)=(\kappa^2-2\ii y\kappa-q)^2$. Comparing the coefficients of $\kappa^3$ and $\kappa$
requires both $p=-4y$ and $p=4yq$, hence $q=-1$, which is impossible. Hence $\Imy L(\kappa_*)\ne0$, and by the symmetry the two
critical values are complex conjugate. The continuation criterion above now applies: all saddles are simple
(section~\ref{sec:saddles}), none reaches a pole or infinity, since the quintic~\eqref{eq:quintic} has the constant
leading coefficient $2$ and takes the value $\mp2\ii r\ne0$ at $\kappa=\pm\ii$, the parameter domain is connected, and the pole orders and decay ends persist. The
indices of the pair are therefore constant. Stokes alignments between the three imaginary saddles can occur; their
indices are determined directly in theorem~\ref{thm:selection}.
\end{proof}

\FloatBarrier
\begin{theorem}[Global selection]\label{thm:selection}
For $r>1$, $p>0$ the three imaginary saddles have zero index and the two off-axis saddles have index of modulus one. The
canonical exponential growth rate is
\begin{equation}
 g(r,p)=\Rey L(\kappa_*;r,p).
 \label{eq:g}
\end{equation}
In the thermal time scale of section~\ref{sec:model}, with $g(r,0)$ defined by continuity, the rate along a moving ray is
\[
 S(v)=\pi^2g\!\left(\frac{\Ra}{\pi^2},\frac{|\Pe-v|}{\pi}\right).
\]
\end{theorem}

\begin{proof}
For the outer imaginary saddles the ascending cycles lie on the imaginary axis. The axis is invariant under
the gradient flow of $\Rey L$, because the symmetry $L(-\overline\kappa)=\overline{L(\kappa)}$ maps it onto itself, and on it
\[
 L(\ii y)=\frac{ry^2}{y^2-1}+y^2-1+py
\]
is real. On each outer interval, $y>1$ and $y<-1$, this function has a single critical point, the saddle (section
\ref{sec:saddles}), and tends to $+\infty$ at both ends, $y\to\pm1$ and $y\to\pm\infty$. The saddle is therefore its strict
minimum, and the two ascending branches run along the axis to the pole and to infinity: to $+\ii$ and $+\ii\infty$ for the
upper saddle, to $-\ii$ and $-\ii\infty$ for the lower one. Neither cycle meets the real contour, and both indices vanish.
For the inner saddle stationarity gives
\begin{equation}
 L(\ii y_0)+1=y_0^2\left[\frac{r(1+y_0^2)}{(1-y_0^2)^2}-1\right]>0 .
 \label{eq:inner}
\end{equation}
Its critical value is real. On the real axis $\Imy L=-p\kappa$, so a cycle with $\Imy L=0$ could meet the axis only at $\kappa=0$,
where $L=-1$; but $\Rey L$ increases along an ascending cycle and cannot pass from a value larger than $-1$ to a smaller one.
The index of the inner saddle is zero as well.

On the real axis $\Imy L=-p\kappa$ takes a given value only once, so each ascending cycle of the pair meets it at most once.
To establish a crossing, fix $r=4$ and let $p\downarrow0$. At $p=0$ the positive temporal maximum is $\kappa=1$, with
$L_{\kappa\kappa}=-4$. The implicit function theorem gives
\[
 \kappa_*(p)=1-\frac{\ii p}{4}+O(p^2),\qquad
 \Imy L(\kappa_*)=-p+O(p^3).
\]
The unique real point with the same value of $\Imy L$ is $x_c=-\Imy L(\kappa_*)/p=1+O(p^2)$, and
\[
 \Rey L(x_c)-\Rey L(\kappa_*)=\frac{p^2}{8}+O(p^3)>0.
\]
The holomorphic Morse lemma therefore places $x_c$ on the local ascending cycle. Its tangent tends to the imaginary
direction, so the crossing is transverse. Reflection gives the crossing for the other saddle. Hence both indices have
modulus one at $r=4$ for small positive $p$, and lemma~\ref{lem:invariance} extends this result to the whole domain.
\end{proof}

Figure~\ref{fig:selection}(b) shows that the height of a critical value does not decide whether a saddle contributes:
two imaginary saddles have positive real critical values on the threshold, but zero index.

\paragraph{Direct quadrature at a fixed probe.}
We evaluated~\eqref{eq:green} at $x=0$ on two lines $\kappa=\xi-\ii c$, with $c=3/4$ and $2/3$.
No pole is crossed, and diffusion makes the connecting vertical integrals vanish at infinity.
The trapezoidal rules use steps $1/64$ and $1/128$ in 55- and 65-digit arithmetic, respectively.
At the $\Pe=50$ threshold and $T=\pi^2t=20$, the peak magnitude of $e^{TL}$ on the shallower line exceeds
$|G(0,t)|$ by a factor of about $1.7\times10^{25}$, so double precision is inadequate.
The two quadratures agree to within $10^{-15}$ of the peak sampled signal in each case.
Neither computation uses the saddle locations.

We fit an exponential sinusoid to $\sqrt{T}\,G(0,t)$ on $16\le T\le20$,
removing the leading algebraic decay before estimating the canonical growth rate and angular frequency.
At point~A of figure~\ref{fig:packet}(a), $(\Pe,\Ra)=(20,95)$, the fit gives $(0.22481,8.27468)$,
against $(0.22488,8.27465)$ for the off-axis pair. The upper imaginary saddle would give a growth rate of $27.42$.
At $\Pe=50$ and $\Ra=\Ra_{\rm a}(50)\simeq208.441$, the fit gives $(-2.5\times10^{-5},22.27621)$,
against $(0,22.27624)$ for the pair, although two imaginary saddles have positive rates $1.79$ and $61.24$.
The differences from the pair predictions decrease when the fitting window is moved later.
The direct signal thus supports the selected growth rates and frequencies in both cases, including neutrality at the threshold.

\subsection{Uniqueness of the threshold}

The selected pair defines a smooth growth function $g(r,p)$. Its derivatives are simple, because the
derivative with respect to the saddle vanishes by stationarity:
\begin{equation}
 g_r=\Rey\frac{\kappa_*^2}{1+\kappa_*^2},\qquad g_p=\Imy\kappa_*=-\eta<0 .
 \label{eq:gderiv}
\end{equation}
The first sign is also fixed, $g_r>0$. To see this, put $q=x_*^2+\eta^2$ and $h=\eta^2$. Reality of $r$ and $p$ in the saddle
condition gives
\begin{equation}
 r=\frac{[(1+q)^2-4h]^2}{(1+q)^2-4qh},\qquad p=\frac{8q\sqrt h\,(1+q-2h)}{(1+q)^2-4qh}.
 \label{eq:qh}
\end{equation}
If $g_r=0$, then $2h=q(1+q)$; since $h<q$ this would imply $q<1$, and substitution into~\eqref{eq:qh} would give
$r=(1-q^2)/(1+2q)<1$, outside the domain. The sign is positive at $p=0$ and hence everywhere.

\begin{theorem}[Absolute threshold]\label{thm:onset}
For every $p>0$ the equation $g(r,p)=0$ has a unique solution $r_0(p)>4$. It is analytic and strictly increasing,
\begin{equation}
 r_0'(p)=\frac{\eta}{g_r}>0,\qquad \Ra_{\rm a}(\Pe)=\pi^2r_0(\Pe/\pi),\qquad \Ra_{\rm a}(0)=4\pi^2 .
 \label{eq:onset}
\end{equation}
The first vertical mode becomes absolutely unstable before all others.
\end{theorem}

\begin{proof}
At $r=4$ the temporal maximum is zero. Shifting the contour to $\kappa=\xi-\ii\epsilon$, $\xi\in\mathbb R$, $0<\epsilon<1$, crosses no pole and gives
$\Rey L\le C\epsilon^2-p\epsilon<0$ with some constant $C$ for small $\epsilon$, so $g(4,p)<0$. For fixed $p$ and $r\to\infty$ the growth
tends to infinity. Positivity of $g_r$ gives existence and uniqueness of the zero, and simplicity of the saddle gives
analyticity. For the vertical wavenumber $\gamma=m\pi$ the threshold is $\Ra_{{\rm a},m}=\gamma^2r_0(\Pe/\gamma)$. On the neutral curve
$rg_r=\Rey(1+\kappa_*^2)+p\eta$, and~\eqref{eq:qh} gives $\Rey(1+\kappa_*^2)=1+q-2h>0$. Therefore
\begin{equation}
 \frac{d}{d\gamma}\bigl[\gamma^2r_0(\Pe/\gamma)\bigr]=\gamma\,\frac{2\Rey(1+\kappa_*^2)+p\eta}{g_r}>0,
\end{equation}
and the threshold increases strictly with the vertical mode number.
\end{proof}

\subsection{The trailing edge stops}\label{sec:edge}

Theorem~\ref{thm:onset} makes the picture of section~\ref{sec:model} precise. For $\Ra>4\pi^2$ the growth $g(r,p)$ is positive at $p=0$, where it equals $r-2\sqrt r$, and decreases strictly in $p$. It also
tends to $-\infty$: shifting the Fourier contour to the line $\Imy\kappa=-1/2$, which lies between the poles, gives
$\Rey L\le C(r)-p/2$ on that line with a constant $C(r)$, and hence $g(r,p)\le C(r)-p/2$. There is therefore a unique finite
speed $c_*(\Ra)>0$, the linear spreading speed of the packet~\cite{VanSaarloos2003}, such that
\begin{equation}
 S(v)>0\quad\Longleftrightarrow\quad |\Pe-v|<c_*(\Ra).
\end{equation}
The edges of the growing packet move with speeds $\Pe\pm c_*$ (figure~\ref{fig:packet}(b)), and the absolute threshold is
the condition $\Pe=c_*(\Ra)$: the trailing edge stops.

The two conjugate contributions of the selected pair add up to a real oscillating response. On the absolute neutral curve
its exponential rate vanishes, but its frequency is nonzero for $p>0$. The edge of the packet is therefore stationary,
while the crests keep moving. For a simple saddle the envelope of the critical signal decays as $t^{-1/2}$, and
exponential growth starts above the threshold.

\subsection{Why the saddle equations need a global selection}\label{sec:heuristics}

\begin{figure}[!htbp]
\centering\includegraphics[width=\textwidth]{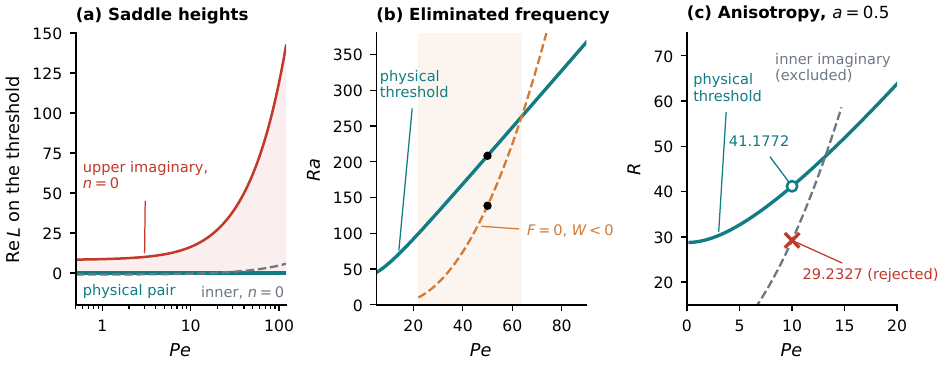}
\caption{Three sources of ambiguity in saddle selection. (a) Growth $\Rey L$ on the physical threshold: zero for the physical
pair, positive for the upper and, at large $\Pe$, the inner imaginary saddles, although their indices vanish. (b) A branch of
the eliminated polynomial $F=0$ of section~\ref{sec:algebraic} without real frequency
($W=\Omega^2<0$) lies below the physical boundary for
$21.9\lesssim\Pe<63.79$; the dots mark $\Pe=50$. (c) Anisotropy, $a=0.5$: physical threshold and the neutral curve of the
excluded inner imaginary saddle; the cross marks the neutral value $R=29.2327$ of
this saddle, shown and rejected in figure~4 of~\cite{CelliBarletta2022}; the circle is the threshold
from~\eqref{eq:universal-onset}. The figure was drawn with the AI-assisted Python code described under \hyperref[sec:computational-methods]{Computational methods}.}
\label{fig:heuristics}
\end{figure}

The neutral-saddle equations have several solutions. Published computations~\cite{BarlettaAlves2017,BarlettaCelli2017,
Barletta2019,AlvesBarletta2015,CelliBarletta2022} decide among them by combining a smallest-threshold search with checks of
contour admissibility, which identify the physical threshold at the parameter values checked. Theorem~\ref{thm:selection}
makes these checks global.
Figure~\ref{fig:heuristics} illustrates three distinct sources of ambiguity.

\emph{Unselected saddle heights} (figure~\ref{fig:heuristics}(a)). On the physical threshold the off-axis pair has $\Rey L=0$, whereas the upper imaginary
saddle has $\Rey L$ between $8$ and $142$ for $0.5\le\Pe\le120$. At larger $\Pe$ the inner imaginary saddle also becomes
positive. Both have zero index. Maximizing $\Rey L$ over all algebraic saddles without checking their contributions would
place the threshold too low.

\emph{Loss of the real-frequency condition under elimination} (figure~\ref{fig:heuristics}(b)). For $21.9\lesssim\Pe<63.79$ the boundary polynomial $F=0$ of section~\ref{sec:algebraic} has a real branch below the
physical boundary on which the squared frequency $W=\Omega^2$ of the neutral wave $\Lambda=\ii\Omega$ is negative. At $\Pe=50$ its root $\Ra=138.6$
($r=14.04$) is smaller than the threshold $208.44$ ($r=21.12$). This root is not a solution of the original neutral-saddle
conditions with real frequency: it appears only after the frequency has been eliminated. The example concerns the algebraic
route, which must be followed by frequency recovery and the condition $W>0$.

\emph{Neutral imaginary saddles below the threshold} (figure~\ref{fig:heuristics}(c)). For anisotropic permeability, $a=0.5$ and $\Pe=10$, the inner imaginary
saddle $k=1.66327\,\ii$ is neutral at $R=29.2327$, well below the physical threshold $41.17721566$ (section~\ref{sec:aniso}).
Celli and Barletta~\cite{CelliBarletta2022} display this solution in their figure~4 and reject it because the steepest-descent
path crosses a singularity. Their admissibility check is essential here: a search for the smallest neutral value without it would return
$29.2327$. By theorem~\ref{thm:selection} the index of this saddle vanishes for all parameters, so the rejection holds
globally.

\section{The exact algebraic boundary}\label{sec:algebraic}

We now eliminate the frequency and obtain the exact curve in the plane of heating and throughflow. The selection theorem is
needed before elimination: it tells which of several algebraic branches to keep.

\subsection{The spurious stationary parabola}

Let $\Delta(\Lambda)$ be the discriminant of the spatial quartic $D$ with respect to $\kappa$. At zero frequency
\begin{equation}
 \Delta(0)=(p^2-4r)^2\bigl[4p^2+(r-4)^2\bigr].
\end{equation}
For $p>0$ the stationary neutral branch is
\begin{equation}
 r=\frac{p^2}{4},\qquad \Ra=\frac{\Pe^2}{4},
 \label{eq:parabola}
\end{equation}
and there $D(\kappa,0)=\bigl(\kappa^2+\tfrac{\ii p}{2}\kappa+1\bigr)^2$. The square factor confirms that both spatial roots are double. Both saddles, however, are imaginary, $\kappa=\ii\bigl[-p\pm\sqrt{p^2+16}\bigr]/4$. For $p>2$ they
are the inner and the lower outer saddles of theorem~\ref{thm:selection}, both of zero index; for $p\le2$ the branch lies in
$r\le1$, where the real-contour estimate already excludes absolute growth. The parabola is therefore not an
absolute-instability boundary. Its discovery and exclusion are due to~\cite{BarlettaAlves2017,BarlettaCelli2017}. The global selection establishes it together with the classification of all other saddles for all parameters.

\subsection{The oscillatory boundary and frequency recovery}

On the physical neutral curve put $\Lambda=\ii\Omega$, $P=p^2/4$, $W=\Omega^2$. The real and imaginary parts of
$\Delta(\ii\Omega)=0$ give two quadratic polynomials $\mathcal U(W)$, $\mathcal V(W)$, whose resultant is
\begin{equation}
 \operatorname{Res}_W(\mathcal U,\mathcal V)=-r^2(r-P)F(r,P).
 \label{eq:resultant}
\end{equation}
The polynomial $F$, given explicitly in appendix~\ref{app:polynomial}, equation~\eqref{eq:F}, has degree seven in $r$
and five in $P$. The extra factor $r-P$ reproduces the spurious parabola. The
equation $F=0$ still contains more than the physical threshold, because eliminating the squared frequency loses the condition $W>0$ (figure~\ref{fig:selection}(c)). The exact physical boundary is therefore
\begin{equation}
 F(r,P)=0,\quad r>4,\quad \mathcal U(W)=\mathcal V(W)=0\ \text{for some } W>0 ,
 \label{eq:exact}
\end{equation}
with $\mathcal U$, $\mathcal V$ and the formula that recovers $W$ also given in
appendix~\ref{app:polynomial}. A positive $W$ restores a real nonzero frequency. The corresponding saddle cannot be imaginary and hence belongs to the unique physical pair, and theorem~\ref{thm:onset}
guarantees uniqueness of the remaining solution. At $\Pe=50$, for example, $F$ has two positive roots with $r>4$:
\begin{center}\small
\begin{tabular}{rrl}\toprule
$r$ & $W$ & meaning\\\midrule
$14.0398794$ & $-2506.1450$ & no real frequency\\
$21.1194932$ & $496.2309$ & $\Ra_{\rm a}=208.4410434$\\\bottomrule
\end{tabular}
\end{center}
The smaller root has no real frequency and is removed by frequency recovery. Choosing the saddle with the largest real part
without checking its index would also give a wrong threshold (section~\ref{sec:heuristics}).

\subsection{The exact crossing and the projection node}

The crossing of the physical boundary with the spurious parabola follows without continuation from
\begin{equation}
 F(r,r)=r(r+4)^4(8r^2-71r+4).
 \label{eq:Frr}
\end{equation}
The only admissible root gives
\begin{equation}
 P_\times=r_\times=\frac{71+17\sqrt{17}}{16},\qquad \Pe_\times=\frac\pi2\sqrt{71+17\sqrt{17}}=18.6583\ldots .
 \label{eq:cross}
\end{equation}
The numerical value of the crossing is known from previous work~\cite{BarlettaAlves2017,BarlettaCelli2017,Barletta2019}.
The factorization~\eqref{eq:Frr} gives its exact form. At this point one pair of physical parameters admits both an oscillatory
neutral state and a non-pinching stationary double root, a feature of the projection that leaves the instability mechanism unchanged.

The plane curve $F=0$ also has a node (figure~\ref{fig:selection}(d)),
\begin{equation}
 (r,P)=(26.65607919,\ 103.06526584),\qquad (\Pe,\Ra)=(63.78756623,\ 263.08495654).
 \label{eq:node}
\end{equation}
The two branches give $W=-5951.9202584$ and $828.8568641$ at the node. Only the latter represents a neutral
oscillation. The physical boundary passes smoothly through the node, as theorem~\ref{thm:onset} requires. Frequency
recovery thus separates a singularity of the projection from a physical encounter of modes.

\section{One growth function for porous convection}\label{sec:universal}

In all generalizations considered here the buoyant response remains rational. Only the position of the poles and the
neutral level of the selected saddle change.

\subsection{General scaling}

Consider the symbol
\begin{equation}
 \lambda(k)=\mathcal R\frac{k^2}{A\pi^2+Bk^2}-(k^2+\pi^2)-\ii\mathcal Pk,\qquad A,B>0,
 \postdisplaypenalty=10000
 \label{eq:general}
\end{equation}
where $\mathcal R$ measures buoyancy and $\mathcal P$ horizontal advection. With
\[
 \alpha=\pi\sqrt{A/B},\qquad \kappa=k/\alpha,\qquad r=\mathcal R/(A\pi^2),\qquad p=\mathcal P/\alpha
\]
one finds
\begin{equation}
 \frac{\lambda}{\alpha^2}=L(\kappa;r,p)-s,\qquad s=\frac BA-1>-1 .
 \label{eq:level-reduction}
\end{equation}
The absolute growth rate in the original time scale is $\alpha^2[g(r,p)-s]$. A constant shift changes neither the saddles
nor their ascending cycles, so the selection theorem transfers to this family and neutrality is $g=s$.

\begin{proposition}[Threshold at a shifted level]\label{prop:level}
For every $s>-1$ and $p\ge0$ the equation $g(r,p)=s$ has a unique solution $r_s(p)>1$. For $p>0$
\begin{equation}
 \frac{\partial r_s}{\partial p}=\frac{\eta}{g_r}>0,\qquad \frac{\partial r_s}{\partial s}=\frac1{g_r}>0,\qquad
 r_s(0)=\bigl(1+\sqrt{1+s}\bigr)^2 .
 \label{eq:level-derivatives}
\end{equation}
Consequently
\begin{equation}
 \mathcal R_{\rm a}(\mathcal P;A,B)=A\pi^2\,r_{B/A-1}\!\left(\frac{\mathcal P}{\pi\sqrt{A/B}}\right).
 \label{eq:universal-onset}
\end{equation}
\end{proposition}

\begin{proof}
On the real contour $\Rey L(\kappa;r,p)=r\kappa^2/(1+\kappa^2)-(1+\kappa^2)\le r-2\sqrt r$ for $r>1$, and the Green function is
bounded by the integral of $e^{t\Rey L}$ along this contour. Hence
\[
 g(r,p)\le r-2\sqrt r\longrightarrow-1\qquad(r\downarrow1),
\]
so $g<s$ near $r=1$ for every $s>-1$; as $r\to\infty$ the growth tends to infinity. Positivity of
$g_r$ gives a single crossing of every level $s>-1$. The derivatives follow from implicit differentiation; at $p=0$ one
solves $r-2\sqrt r=s$.
\end{proof}

For the vertical mode $m$,
\begin{equation}
 \mathcal R_{{\rm a},m}(\mathcal P)=m^2\mathcal R_{{\rm a},1}(\mathcal P/m),
 \label{eq:vertical-scaling}
\end{equation}
so the whole sequence of vertical thresholds is expressed through one curve.

Physically, the class comes from the linearized drag. Consider transverse-roll disturbances, whose roll axes are
perpendicular to the basic flow, of a uniform throughflow $U\mathbf e_x$ with $U=\Pe>0$ in the units of
section~\ref{sec:model}, under an instantaneous isotropic drag law $d(|\mathbf u|)\,\mathbf u$. Linearization gives the drag
tensor $d_\perp I+(d_\parallel-d_\perp)\,\mathbf e_x\mathbf e_x^{\mathsf T}$ with $d_\perp=d(U)$ and $d_\parallel=d(U)+U\,d'(U)$. A
disturbance along the flow changes both the speed and the drag coefficient, whereas one across the flow only turns the
velocity. These disturbances obey the symbol~\eqref{eq:general} with $A=d_\parallel$ and $B=d_\perp$, up to a common factor
absorbed into $\mathcal R$. Forchheimer drag, $d=1+\mathit{Fo}\,|\mathbf u|$ with the Forchheimer number $\mathit{Fo}$, gives
$A=1+2q$ and $B=1+q$, where $q=\mathit{Fo}\,U$. A power-law fluid with index $\nu$, $d\propto|\mathbf u|^{\nu-1}$, gives
$A=\nu$ and $B=1$ after the common factor $d(U)$ is absorbed into $\mathcal R$. Anisotropic permeability gives
$B/A=K_h/K_z$ directly. Every instantaneous isotropic drag law with $d_\parallel,d_\perp>0$ thus leads to a level of the
same growth function, while the open upper boundary changes the admissible vertical wavenumbers. The statement concerns
the linearized problem for these disturbances, not the nonlinear models.

The level also has a direct mechanical meaning:
\[
 1+s=\frac{d(U)}{[U\,d(U)]'},\qquad
 s=-\frac{U\,d'(U)}{d(U)+U\,d'(U)}.
\]
The numerator in the first ratio is the secant resistance of the pressure--flow curve, and the denominator is its tangent
resistance. Thus $d_\parallel>0$ expresses local monotonicity of this curve. Equivalently, its local logarithmic slope
$\nu_{\rm eff}=\mathrm d\log[U\,d(U)]/\mathrm d\log U$ gives $s=\nu_{\rm eff}^{-1}-1$. Under the assumed positive resistances,
decreasing $d(U)$ gives $s>0$, while increasing $d(U)$ gives $s<0$. The physical threshold also depends on the resistance
scale and on the rescaled advection parameter.

\paragraph{Interphase heat exchange.}
The same level construction applies to a thermal sink. In the high-conductivity limit for the solid matrix
of Freitas \emph{et al.}~\cite[Eq.~(16)]{Freitas2022}, the dispersion relation for rolls uniform in the
spanwise direction is, with $\mu_m=m\pi$,
\[
 \lambda_m(k)=R\frac{k^2}{k^2+\mu_m^2}-(k^2+\mu_m^2)-H-\ii\Pe k .
\]
The heat exchange term $-H$ shifts the growth rate without changing the saddle locations or the contours at fixed
$R$ and $\Pe$. Hence
\[
 R_{{\rm a},m}(\Pe,H)=\mu_m^2 r_{H/\mu_m^2}(\Pe/\mu_m),\qquad
 \left.\frac{\partial R_{{\rm a},m}}{\partial H}\right|_{\Pe,m}=\frac1{g_r}>0
\]
by proposition~\ref{prop:level}. At $\Pe=10$ and $H=30$, this gives $R_{{\rm a},1}=102.28205$, within
$6\times10^{-5}$ of the value $102.2820$ reported for the spanwise-uniform mode in figure~8 of~\cite{Freitas2022}.

\subsection{The universality map}

Thresholds can then be computed without any search over complex wavenumbers.

\newcounter{algorithm}
\newcommand{\algobox}[2]{\par\medskip\noindent\refstepcounter{algorithm}\label{#1}%
\fbox{\parbox{\dimexpr\linewidth-2\fboxsep-2\fboxrule}{\textbf{Algorithm~\thealgorithm.} #2}}\par\medskip}
\algobox{alg:onset}{\emph{Absolute threshold of the first vertical mode for any model of the form~\eqref{eq:general}.}
\begin{enumerate}[leftmargin=*,itemsep=1pt,topsep=2pt]
\item From the model parameters form $A$, $B$ and $\mathcal P$; set $s=B/A-1$ and
$p=\mathcal P/(\pi\sqrt{A/B})$.
\item Bracket the root: by proposition~\ref{prop:level}, $g(r,p)-s<0$ as $r\to1^+$ and $g\to\infty$ as $r\to\infty$;
start from $r_{\rm hi}=8$ or any larger value and double $r_{\rm hi}$ until $g(r_{\rm hi},p)>s$.
\item Solve $g(r,p)=s$ on $(1,r_{\rm hi})$ by a bracketing root finder, which keeps a sign change while shrinking the
interval: bisection, or Brent's method~\cite{Brent1971}, used in the implementation. Each evaluation of $g$ takes
the off-axis root of the quintic~\eqref{eq:quintic} with positive real part. By proposition~\ref{prop:level} the root
is unique.
\item Return $\mathcal R_{\rm a}=A\pi^2 r_s(p)$, or use the explicit parametrization of section~\ref{sec:parametrization}
instead of steps 2--3.
\end{enumerate}
Other vertical modes follow from~\eqref{eq:vertical-scaling}. The selection is done once by theorem~\ref{thm:selection}, so a single scalar equation with a guaranteed bracket replaces
the three real nonlinear equations solved at each parameter point.}

\begin{figure}[!htbp]
\centering\includegraphics[width=\textwidth]{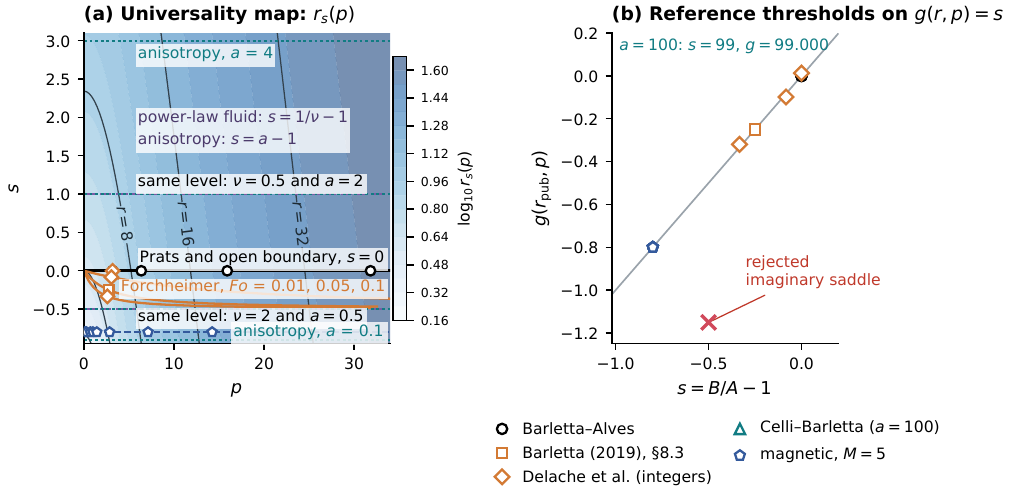}
\caption{Universality. (a) The threshold $r_s(p)$ on the plane $(p,s)$ (color and contours) with the model trajectories:
Prats and open boundary ($s=0$), Forchheimer drag with $q=\mathit{Fo}\,\Pe$, $\mathit{Fo}=0.01$, $0.05$, $0.1$, power-law fluid ($s=1/\nu-1$) and
anisotropy ($s=a-1$), and magnetic resistance ($M=5$, $s=-0.8$); markers are tabulated values.
(b) Reference thresholds mapped to $(r,p,s)$
by~\eqref{eq:universal-onset}: Barletta and Alves~\cite{BarlettaAlves2017}, the monograph~\cite{Barletta2019} (\S8.3),
Delache \emph{et al.}~\cite{Delache2007} (integers quoted in their section~4.1) and Celli and
Barletta~\cite{CelliBarletta2022} ($a=100$, $s=99$, given as a number), and the magnetic case $M=5$ of
Turkyilmazoglu~\cite{Turkyilmazoglu2023}. The cross is the neutral value $29.2327$ of the inner imaginary
saddle, shown and rejected in figure~4 of~\cite{CelliBarletta2022}. The figure was drawn with the AI-assisted Python code described under \hyperref[sec:computational-methods]{Computational methods}.}
\label{fig:universality}
\end{figure}

Figure~\ref{fig:universality}(a) shows $r_s(p)$ on the plane $(p,s)$ with representative model trajectories.
The Prats problem and the layer with an open upper boundary lie on $s=0$. Forchheimer drag with $q=\mathit{Fo}\,\Pe$ (Forchheimer number $\mathit{Fo}$, section~\ref{sec:forch}) moves from $s=0$
towards $s=-1/2$. Transverse rolls of a power-law fluid with index $\nu$ and anisotropic permeability with ratio
$a=K_h/K_z$ (sections~\ref{sec:powerlaw} and~\ref{sec:aniso}) give the horizontal levels $s=1/\nu-1$ and $s=a-1$. Coincidences of levels are immediate: a power-law fluid with $\nu=2$ and anisotropy with $a=1/2$ share
$s=-1/2$, and $\nu=1/2$ and $a=2$ share $s=1$. Their thresholds differ only by the scales in~\eqref{eq:universal-onset}.
A vertical magnetic field with resistance factor $M$ gives the level $s=1/M-1$ (section~\ref{sec:magnetic}).
Figure~\ref{fig:universality}(b) tests the tabulated transverse-mode benchmarks, mapped to $(r,p,s)$ and inserted in $g$.
The tabulated magnetic and zero-field thresholds agree with the calculation to better than $10^{-6}$ relative error. The Forchheimer thresholds quoted as integers by Delache \emph{et al.} agree to within $0.5\%$, two of the three to their quoted precision. The rejected imaginary-saddle value $29.2327$ shown in
figure~4 of~\cite{CelliBarletta2022} falls off the diagonal, $g=-1.151$ at $s=-1/2$.

\FloatBarrier
\subsection{Forchheimer drag}\label{sec:forch}

In the model of Delache \emph{et al.}~\cite{Delache2007} the inertial drag is characterized by $q=\mathit{Re}_K=\mathit{Fo}\,\Pe$, where $\mathit{Fo}$ is the
Forchheimer number (denoted $F$ in~\cite{Delache2007}). For
transverse rolls their dispersion relation (equation~(9) of~\cite{Delache2007} with $m=0$) has the buoyant denominator
$(1+2q)\pi^2+(1+q)k^2$, so $A=1+2q$, $B=1+q$, $s=-q/(1+2q)$ and
\begin{equation}
 \Ra_{\rm a}=\pi^2(1+2q)\,r_{-q/(1+2q)}\!\left(\frac{\Pe}{\pi}\sqrt{\frac{1+q}{1+2q}}\right).
 \label{eq:forchheimer}
\end{equation}
Under a change of observer only $\Pe$ in the argument is replaced by $|\Pe-v|$. The drag parameter $q$ is set by the basic
flow and stays fixed. The temporal threshold is
\begin{equation}
 \Ra_T=\pi^2\bigl(\sqrt{1+2q}+\sqrt{1+q}\bigr)^2 .
\end{equation}
At $\Pe=10$ formula~\eqref{eq:forchheimer} gives $57.80$, $66.29$ and $141.31$ for $\mathit{Fo}=0$, $0.01$, $0.1$. Delache \emph{et al.}
quote these thresholds as the integers $58$, $66$ and $142$ (section~4.1 of~\cite{Delache2007}; also p.~56 of~\cite{Delache2005}). The first two agree after rounding to the nearest integer, while $142$ exceeds $141.31$ by $0.49\%$, a difference whose origin cannot be determined from the paper. For Barletta's form-drag parameter $G=0.05$ at $\Pe=10$, that is $q=G\Pe=0.5$, formula~\eqref{eq:forchheimer} gives
$99.7984$, the value in equation~(8.69) of~\cite{Barletta2019}.

\subsection{Transverse rolls of a power-law fluid}\label{sec:powerlaw}

In the setting of Alves and Barletta~\cite{AlvesBarletta2015} let $\nu>0$ be the power-law index. For transverse rolls $\mathcal R=R/\Pe^{\nu-1}$, where $R$ is their Darcy--Rayleigh number, and $A=\nu$, $B=1$;
hence
\begin{equation}
 R_{\rm a}=\nu\pi^2\Pe^{\nu-1}\,r_{(1-\nu)/\nu}\!\left(\frac{\Pe}{\pi\sqrt\nu}\right),\qquad
 R_T=\pi^2\Pe^{\nu-1}(1+\sqrt\nu)^2 .
 \label{eq:powerlaw}
\end{equation}
These formulas describe the family of transverse disturbances. The competition with longitudinal and oblique rolls studied
in~\cite{AlvesBarletta2015} requires a comparison of the corresponding families and is outside the present class. Within the
transverse problem the whole complex selection is already done by theorem~\ref{thm:selection}.

\subsection{Anisotropic permeability}\label{sec:aniso}

\begin{figure}[!htbp]
\centering\includegraphics[width=\textwidth]{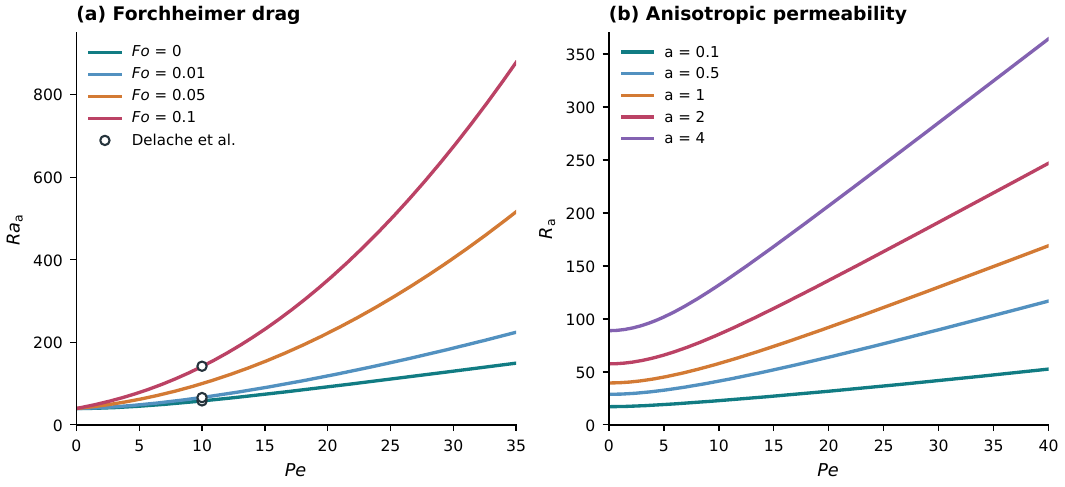}
\caption{Two applications in physical variables. (a) Forchheimer drag with $q=\mathit{Fo}\,\Pe$; markers are rounded published values at
$\Pe=10$. (b) Anisotropic permeability for several $a=K_h/K_z$. All curves follow from~\eqref{eq:universal-onset} without any
optimization over the complex wavenumber. The figure was drawn with the AI-assisted Python code described under \hyperref[sec:computational-methods]{Computational methods}.}
\label{fig:models}
\end{figure}

Let $a=K_h/K_z$ be the ratio of horizontal to vertical permeability. In the normalization of Celli and
Barletta~\cite{CelliBarletta2022} the Rayleigh number is built on $K_h$, and $A=1$, $B=a$, $s=a-1$. Therefore
\begin{equation}
 R_{\rm a}=\pi^2\,r_{a-1}\!\left(\frac{\Pe\sqrt a}{\pi}\right).
 \label{eq:anisotropy}
\end{equation}
Increasing $a$ raises both the argument and the level, and both sensitivities in~\eqref{eq:level-derivatives} are positive:
\begin{equation}
 \frac{\partial R_{\rm a}}{\partial\Pe}=\frac{\pi\sqrt a\,\eta}{g_r}>0,\qquad
 \frac{\partial R_{\rm a}}{\partial a}=\frac{\pi^2}{g_r}\left(1+\frac{\Pe\,\eta}{2\pi\sqrt a}\right)>0 .
\end{equation}
The monotonicities found numerically in~\cite{CelliBarletta2022} are thus confirmed analytically. At $a=100$, $\Pe=10$ one
obtains the published threshold $1545.47$; at $a=0.5$, $\Pe=10$ the threshold is $41.1772$
(section~\ref{sec:heuristics}). At $\Pe=0$, $R_{\rm a}=\pi^2(1+\sqrt a)^2$, and for fixed $\Pe$ and $a\to0$, $R_{\rm a}\to\pi^2$, as found
in~\cite{CelliBarletta2022}.
The last limit needs no expansion near the degenerate level $s=-1$: for every fixed $r>1$, $g(r,0)=-1+(\sqrt r-1)^2>-1$, so for
small $a$ the threshold lies between $1$ and any such $r$. Figure~\ref{fig:models} compares anisotropy with inertial drag in
physical variables.

\subsection{An open upper boundary}

In the problem of Barletta and Celli~\cite{BarlettaCelli2017} the vertical wavenumbers are $(m-\tfrac12)\pi$. For the first
mode
\begin{equation}
 \Ra_{\rm a}^{\rm BC}(\Pe)=\tfrac14\Ra_{\rm a}(2\Pe),
 \label{eq:open-boundary}
\end{equation}
an equivalence already stated in their equation~(35). Together with the selection theorem it transfers the whole exact
boundary, the expansions and the convergence certificates to the open layer. Both radii in the variable $\Pe$ are halved.

\subsection{A vertical magnetic field}\label{sec:magnetic}

Capone and Rionero~\cite{CaponeRionero2016} found magnetic stabilization in a modal and energy stability analysis of
porous convection. Here the threshold of interest is absolute instability at prescribed horizontal throughflow.
In Turkyilmazoglu's Darcy model, the Lorentz force adds a dimensionless resistance $Q\geq 0$ to the horizontal resistance
and leaves the vertical resistance unchanged. His dispersion relation for transverse rolls~\cite[Eq.~(12)]{Turkyilmazoglu2023}
has the form~\eqref{eq:general} with $A=M=1+Q$, $B=1$, $\mathcal R=\Ra$ and
$\mathcal P=\Pe$. Hence the first-mode threshold is
\[
 \Ra_{\rm a}(\Pe,Q)=M\pi^2 r_{1/M-1}\left(\frac{\Pe}{\pi\sqrt M}\right).
\]
Tables~1 and~2 of~\cite{Turkyilmazoglu2023} give $19$ finite-$\Pe$ benchmarks at zero field and $M=5$.
For example, at $\Pe=10$ the calculation
gives $\Ra_{\rm a}=57.80357$ for $M=1$ and $142.34085$ for $M=5$. The leading large-$\Pe$ term
$c_1\sqrt M\,\Pe$ and the limiting saddle $k_\infty=\pi\sqrt M\,\kappa_\infty$ agree with
Eq.~(24) of~\cite{Turkyilmazoglu2023}; the second term follows from the general expansion
in section~\ref{sec:resistance-asymptotic}.

\paragraph{Monotonicity at prescribed throughflow.}
For the entire class~\eqref{eq:general}, $\mathcal R_{\rm a}(\mathcal P;A,B)$ increases strictly
with either $A$ or $B$ when the other coefficient and $\mathcal P$ are held fixed.
In particular, increasing $Q$ at fixed $\Pe$ raises the absolute threshold.

\begin{proof}
It suffices by reflection to take $\mathcal P>0$. Let $w=\kappa_*^2$ at the selected saddle on the
threshold. The contribution from the motion of the saddle vanishes upon differentiating
$\Rey\lambda(k_*)=0$, since $\lambda_k(k_*)=0$. At fixed throughflow the derivatives are therefore
\[
 \frac{A}{\mathcal R_{\rm a}}\,\partial_A\mathcal R_{\rm a}
 =\frac{\Rey[w/(1+w)^2]}{g_r},\qquad
 \frac{B}{\mathcal R_{\rm a}}\,\partial_B\mathcal R_{\rm a}
 =\frac{\Rey[(w/(1+w))^2]}{g_r},\qquad
 g_r=\Rey\frac{w}{1+w}>0 .
\]
To determine the signs, put $q=|w|$, $\delta=\Rey w$ and $\eta=-\Imy\kappa_*>0$.
The stationarity identities~\eqref{eq:qh} read
\[
 D_q=1+2q(1+\delta)-q^2,\qquad
 r=\frac{(1+2\delta+q^2)^2}{D_q},\qquad p=\frac{8\eta q(1+\delta)}{D_q}.
\]
Since $r,p>0$, $D_q>0$ and $\delta>-1$. Neutrality, $rg_r-\delta-p\eta=B/A$, then gives
\[
 \frac{B}{A}D_q=q^4-3q^2+2(1+q)\delta^2+2q\delta .
\]
For $0<q<1$, the right-hand side is a convex quadratic in $\delta$. Its values at $\delta=-q$ and $\delta=0$
are $q^2(q-1)(q+3)$ and $q^2(q^2-3)$, respectively, so it is negative throughout $[-q,0]$. Neutrality
therefore requires $\delta>0$ for $q<1$. Now $|1+w|^4\Rey[w/(1+w)^2]=(1+q^2)\delta+2q^2$.
This is positive for $q<1$ and exceeds $q^2-1$ for $q\ge1$ since $\delta>-1$.
For the other derivative, stationarity gives $rw/(1+w)^2=w+\ii p\kappa_*/2$.
Subtracting its real part, $\delta+p\eta/2$, from $rg_r$ and using neutrality gives
$r\,\Rey[(w/(1+w))^2]=B/A+p\eta/2>0$.
At $\mathcal P=0$, $\mathcal R_{\rm a}=\pi^2(\sqrt A+\sqrt B)^2$, which has the same
strict monotonicities.
\end{proof}

Figure~6(d) of Turkyilmazoglu~\cite{Turkyilmazoglu2023} compares thresholds for different $M$ by plotting $\Ra_{\rm a}$ against
$\Imy k_*$. The curves cross in this plot, where points at the same $\Imy k_*$ generally have different imposed throughflows. At $\Imy k_*=-3.2$, for example,
the pairs $(\Pe,\Ra_{\rm a})$ are approximately $(87.96,359.19)$ at $M=1$ and $(12.93,160.18)$ at $M=5$.
The lower threshold in the second pair does not describe a change of magnetic field alone.
At fixed $\Pe$ the order is instead fixed by the monotonicity just proved.

\subsection{Explicit parametrization of every level}\label{sec:parametrization}

Write $q=|\kappa_*|^2$, $h=(\Imy\kappa_*)^2$ and $B_q=(1+q)^2-4qh$. Substituting~\eqref{eq:qh} into $g(r,p)=s$
gives a quadratic in $h$:
\begin{equation}
 N_s(q,h)=8(q+1)h^2-4q(2q-s+2)h+q^4+2q^3-sq^2-2sq-2q-s-1=0 .
 \label{eq:Ns}
\end{equation}
This is the neutral-saddle equation of Turkyilmazoglu~\cite[Eq.~(18)]{Turkyilmazoglu2023} in different variables.
For $k=k_r+\ii k_i$, substituting $k_r^2=M\pi^2(q-h)$, $k_i^2=M\pi^2h$ and $s=1/M-1$ gives his numerator as
$(M\pi^2)^4N_s(q,h)$. His radical parametrization uses $k_i$ and covers the Prats and magnetic cases, $-1<s\leq0$.
The quadratic form uses $q$ and works for every $s>-1$, with the contributing saddle specified by
theorem~\ref{thm:selection}. The two roots at fixed $q$ are
\begin{equation}
 h(q)=\frac{q(2q-s+2)\pm\sqrt{-T_s(q)}}{4(q+1)},
 \label{eq:level-radical}
\end{equation}
with
\begin{equation}
 T_s(q)=2q^5+2q^4+(2s-4)q^3-(s^2+2s+8)q^2-6(s+1)q-2(s+1).
 \label{eq:Ts}
\end{equation}
Once $h$ is chosen, heating, throughflow and the angular frequency $\Omega$ (in canonical units, with the phase convention
$e^{\lambda t+\ii kx}$) follow:
\begin{equation}
 r=\frac{[(1+q)^2-4h]^2}{B_q},\qquad p=\frac{8q\sqrt h\,(1+q-2h)}{B_q},\qquad
 \Omega=-\frac{8\sqrt h\,(1+q)(q-h)^{3/2}}{B_q}.
 \label{eq:frequency}
\end{equation}
The physical arc starts at $q=\sqrt{1+s}$, $h=0$ and is continued with $0<h<q$, $B_q>0$ and $p>0$. The sign of the radical
in~\eqref{eq:level-radical} changes where the two branches meet, at a zero of $T_s$. The selection theorem guarantees that every $p$ corresponds to exactly
one point of this arc. For a fixed anisotropy ratio this gives the curves of
figure~\ref{fig:models}(b) directly, since the level $s=a-1$ is constant along each of them. For the saddle with positive real part $\Omega<0$, so the phase moves downstream.

\section{Elliptic geometry and expansions of the threshold}\label{sec:geometry}

\begin{figure}[!t]
\centering\includegraphics[width=\textwidth]{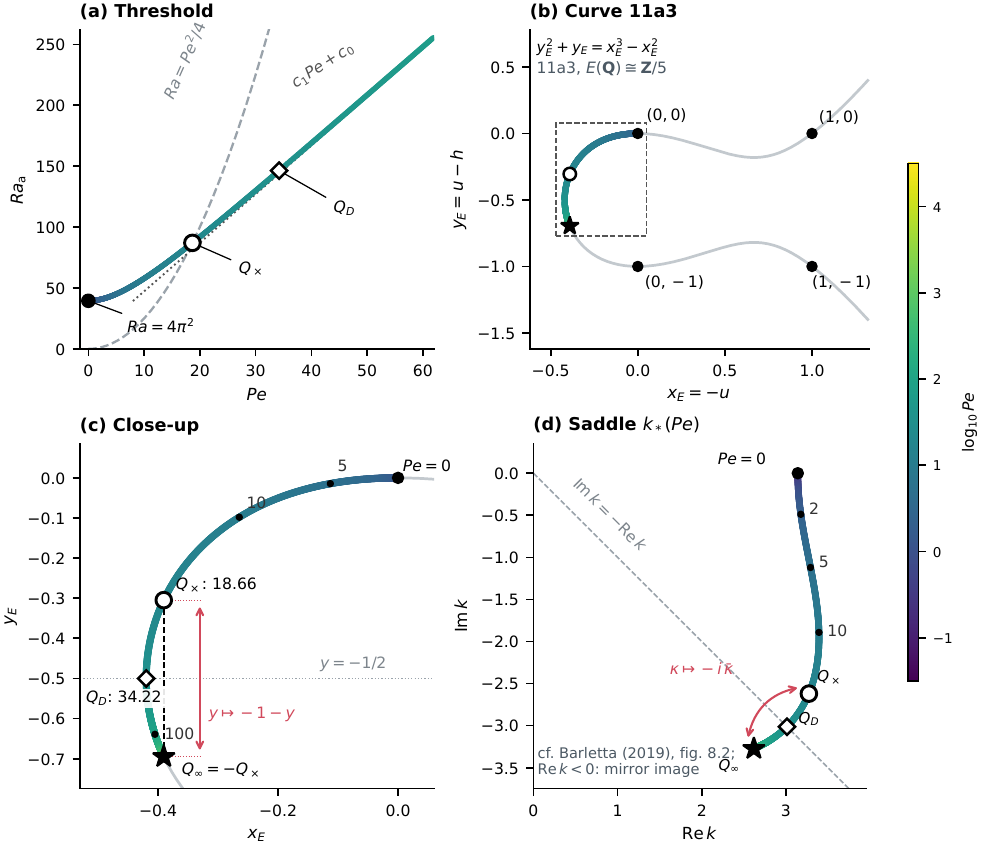}
\caption{One physical arc in four charts. (a) Threshold $\Ra_{\rm a}(\Pe)$ and its crossing $Q_\times$ with the parabola
$\Ra=\Pe^2/4$ at $\Pe=18.658$; the dotted line is the two-term asymptote $c_1\Pe+c_0$ of~\eqref{eq:large}. (b) The curve 11a3, a model of the modular curve $X_1(11)$, with its rational points and the physical arc, colored by
$\log_{10}\Pe$; the four marked rational points and $\mathcal O$ at infinity form $E(\mathbb Q)\cong\mathbb Z/5$~\cite{LMFDB11a3}. (c) Close-up of the arc: $Q_\times$ and the limit point $Q_\infty$ ($\Pe\to\infty$)
are opposite in the group law. The point $Q_D$ ($\Pe=34.22$) is the vertical tangent of the elliptic curve; the physical threshold in (a) is smooth there.
(d) The same arc in the chart $k=\pi\kappa$, that is, the trajectory of $k_*=\pi\kappa_*$, comparable with figure~8.2
of~\cite{Barletta2019}. The reflection
$y_E\mapsto-1-y_E$ corresponds to $\kappa\mapsto-\ii\bar\kappa$ in the diagonal $\Imy k=-\Rey k$. The figure was drawn with the AI-assisted Python code described under \hyperref[sec:computational-methods]{Computational methods}.}
\label{fig:elliptic}
\end{figure}

The physical boundary is one arc of a known elliptic curve (figure~\ref{fig:elliptic}(a,b)). Parametrizing this arc computes
the threshold across the whole range of throughflow, and its two end points generate two different convergent series.

\subsection{One arc instead of saddle continuation}

For $s=0$ replace $q$ by $1+2u$. After removing the factor $q+1$, equation~\eqref{eq:Ns} becomes
\begin{equation}
 E(u,h)=h^2-(1+2u)h+u(1+u)^2=0 .
 \label{eq:E}
\end{equation}
Equation~\eqref{eq:E} describes the off-axis oscillatory neutral locus. Its physical arc gives the absolute-instability threshold. The stationary imaginary saddles form the separate parabola~\eqref{eq:parabola}.

With $\mathcal A=(1+u)^2-h$, $\mathcal B=(1+u)^2-(1+2u)h$ and $\mathcal C=(1+2u)^2h(1+u-h)^2$,
\begin{equation}
 r=\frac{4\mathcal A^2}{\mathcal B},\qquad P=\frac{4\mathcal C}{\mathcal B^2},\qquad \Pe=2\pi\sqrt P,\qquad \Ra=\pi^2r ,
 \label{eq:elliptic-map}
\end{equation}
and $h$ is found from one square root,
\begin{equation}
 h_\pm(u)=\frac{1+2u\pm\sqrt{1-4u^2-4u^3}}2 .
 \label{eq:radical}
\end{equation}
The full route has two parts. First take $h_-$ and increase $u$ from zero to the positive root of $4u_D^3+4u_D^2-1=0$,
$u_D=0.419643\ldots$; then take $h_+$ and decrease $u$ to
\begin{equation}
 u_\infty=\frac{\sqrt{17}-1}{8},\qquad h_\infty=\frac{5+3\sqrt{17}}{16}.
 \label{eq:infinity}
\end{equation}
At the start $\Pe=0$ and $\Ra=4\pi^2$; at the end $\mathcal B\to0^+$ and $\Pe\to\infty$. The change of sign of the radical
occurs at $\Pe\simeq34.216$, where the physical threshold remains smooth. The whole range of throughflows is thus computed
without an initial guess for a complex root.

The change of coordinates $x_E=-u$, $y_E=u-h$ brings~\eqref{eq:E} to
\begin{equation}
 y_E^2+y_E=x_E^3-x_E^2 ,
 \label{eq:weierstrass}
\end{equation}
the curve 11a3 in Cremona's notation (11.a3 in the LMFDB~\cite{LMFDB11a3}), with discriminant $-11$ and
$j$-invariant $-4096/11$. It is a model of the modular curve $X_1(11)$~\cite{LMFDB11a3}. In the variables $q=|\kappa|^2$, $h=(\Imy\kappa)^2$ equation~\eqref{eq:E} is the polar curve $|\kappa|^6+|\kappa|^4\cos4\varphi-|\kappa|^2=1$, $\varphi=\arg\kappa$. Barletta and Celli~\cite{BarlettaCelli2017} use it to describe the off-axis part of the set of
neutral saddles; the rest of that set is the imaginary axis. The physical arc is the saddle trajectory drawn in figure~8.2 of~\cite{Barletta2019};
Turkyilmazoglu~\cite{Turkyilmazoglu2023} also parametrizes the nonmagnetic locus in radicals. In the present
description $r$ and $P$ are rational functions of the coordinates of the elliptic model.
The two square-root branches describe one selected arc, including its passage through the
turning point of the coordinate $u$.

The crossing with the spurious parabola has the same $u=u_\infty$ but
\begin{equation}
 h_\times=\frac{7+\sqrt{17}}{16},\qquad y_{E,\times}+y_{E,\infty}=-1 ,
\end{equation}
so on~\eqref{eq:weierstrass} the two points are opposite in the group law. Their complex wavenumbers satisfy
$\kappa_\times=-\ii\,\overline{\kappa_\infty}$: in the saddle chart the negation is the reflection in the diagonal
$\Imy k=-\Rey k$, and the point $Q_D$ where the radical changes sign is its fixed point (figure~\ref{fig:elliptic}c,d). This symmetry links the finite crossing with the limit of strong throughflow and explains why $\sqrt{17}$ appears in both.

\subsection{Why a general level has genus two}

Completing the square in~\eqref{eq:Ns} gives
\begin{equation}
 y^2=-T_s(q),\qquad y=4(q+1)h-q(2q-s+2).
\end{equation}
The discriminant of the quintic is
\begin{equation}
 \operatorname{Disc}_qT_s=16s^4(s+1)(s^2+3s+88)(54s^4+239s^3+384s^2-80s+64).
\end{equation}
The last two factors are positive on the real axis. For the quartic this is seen directly: for $s\ge0$ the part
$384s^2-80s+64$ is positive and the other terms are non-negative, while for $s<0$ one has $s^2(54s^2+239s+384)-80s+64>0$, because the quadratic $54s^2+239s+384$ has negative
discriminant and $-80s>0$. In the physical domain $s>-1$ the
quintic therefore has distinct roots for $s\ne0$. The map $(q,y)\mapsto q$ from the curve $y^2=-T_s(q)$ to the Riemann sphere has degree $n=2$ and six simple branch
points: the five roots of $T_s$ and $q=\infty$, where the odd degree of $T_s$ makes $y\sim q^{5/2}$ change sign around
infinity. The Riemann--Hurwitz formula $2g-2=n(2g_0-2)+\sum_p(e_p-1)$~\cite{Hartshorne1977}, with genus $g_0=0$ of the
sphere and ramification index $e_p=2$ at each branch point, gives $2g-2=2\cdot(-2)+6$, that is, genus $g=2$. At $s=0$ the polynomial~\eqref{eq:Ns} factorizes:
\begin{equation}
 N_0=(q+1)\bigl[8h^2-8qh+(q-1)(q+1)^2\bigr],
\end{equation}
and the second factor is the elliptic curve~\eqref{eq:E}. The elliptic case is a degeneration of a genus-two family, and the global selection applies beyond it. The group law, the rational torsion point at $\Pe=0$ and the elliptic inversion formula below
are specific to $s=0$, that is, to the Prats problem and the equivalent open-boundary problem.

\FloatBarrier
\subsection{Weak throughflow and integer coefficients}

For a weak throughflow the threshold is naturally compared with $4\pi^2$. With $z=\Pe^2/(16\pi^2)$ and
$Y(z)=\Ra_{\rm a}(\Pe)/(4\pi^2)$,
\begin{equation}
 Y(z)=1+z-z^2+3z^3-13z^4+68z^5-400z^6+2549z^7-17211z^8+\cdots .
 \label{eq:small}
\end{equation}
The leading shift $\Ra_{\rm a}-4\pi^2\approx\Pe^2/4$ follows from the Ginzburg--Landau approximation in Dufour and
N\'eel~\cite[Sec.~III B]{DufourNeel1998} and Chung \emph{et al.}~\cite[Eqs.~(21) and~(22)]{ChungKimChoi2009};
see also~\cite{Deissler1987,Delache2005}.
The exact expansion differs from this approximation starting at order $\Pe^4$. Odd powers of $\Pe$ are absent because the threshold does not
depend on the direction of the flow.

The coefficients follow directly from the boundary polynomial of appendix~\ref{app:polynomial}. Since $r=4Y$ and $P=4z$, the
threshold satisfies
\[
 F\bigl(4Y(z),4z\bigr)=0,\qquad Y(0)=1 .
\]
Substituting $Y=1+Y_1z+Y_2z^2+\cdots$ and equating the coefficients of equal powers of $z$ gives, after division by
$4F_r(4,0)=200704$, where $F_r=\partial F/\partial r$, one linear equation for each new coefficient. With previously determined coefficients substituted
at each step, the first three orders give $Y_1=1$, $Y_2=-1$ and $Y_3=3$. In general, if $Y_1,\dots,Y_{n-1}$ are known,
\begin{equation}
 Y_n=-\frac1{200704}\,[z^n]\,F\Bigl(4\Bigl[1+\sum_{j=1}^{n-1}Y_jz^j\Bigr],\,4z\Bigr),
 \postdisplaypenalty=10000
 \label{eq:small-recursion}
\end{equation}
where $[z^n]$ denotes the coefficient of $z^n$ in the polynomial that follows it. Because $F_r(4,0)=50176\ne0$, every coefficient is determined uniquely. The
recursion produces rational numbers; that they are integers is a separate statement.

Integrality of all coefficients is an exact property of this normalization. From~\eqref{eq:elliptic-map}, on the branch
$u(0)=h(0)=0$,
\[
 E(u,h)=0,\qquad z\mathcal B^2=\mathcal C,\qquad Y=\mathcal A^2/\mathcal B .
\] At $z=u=h=0$ the Jacobian of the first two equations
with respect to $(u,h)$ is $\bigl(\begin{smallmatrix}1&-1\\0&-1\end{smallmatrix}\bigr)$, with determinant $-1$. Determining the
coefficients order by order, each step solves the same integer system with determinant $-1$, so $u,h\in\mathbb Z[[z]]$. The first terms are
$u=z-3z^2+14z^3-83z^4+\cdots$ and $h=z-2z^2+9z^3-54z^4+\cdots$. Since $\mathcal B(0)=1$, division preserves
integrality, and
\begin{equation}
 Y\in\mathbb Z[[z]].
\end{equation}
By Eisenstein's theorem~\cite{Eisenstein1852} the coefficients of any algebraic series with rational coefficients become
integral after a suitable integer rescaling of the argument and the value. Here the natural physical normalization already does this. At other rational levels with rational $\sqrt{1+s}$ a nontrivial rescaling may be needed. In practice any order of the small series follows from an exact integer recursion, without repeated differentiation of a complex saddle. Both recursions,~\eqref{eq:small-recursion} and the integer lifting of $(u,h)$, are implemented in the
script \texttt{prats\_series.py}, which checks that they agree. At the time of writing, the OEIS contains no matching entry for the signed coefficients of~\eqref{eq:small} or their
absolute values~\cite{OEISPrats}.

\subsection{Strong throughflow}

As $\Pe\to\infty$ the physical saddle tends to
\begin{equation}
 \kappa_\infty=\frac{\sqrt{7+\sqrt{17}}}{4}-\ii\frac{\sqrt{5+3\sqrt{17}}}{4} ,
 \label{eq:kinfinity}
\end{equation}
a formula that, together with the leading coefficient $c_1$, is known from
\cite{BarlettaCelli2017,Barletta2019,Turkyilmazoglu2023}. We continue
the expansion in closed form,
\begin{equation}
 \Ra_{\rm a}=c_1\Pe+c_0+\frac{c_{-1}}{\Pe}+\frac{c_{-2}}{\Pe^2}+O(\Pe^{-3}),
 \label{eq:large}
\end{equation}
with
\begin{equation}
\begin{gathered}
 c_1=\frac\pi8\sqrt{51\sqrt{17}-107},\qquad c_0=\frac{\pi^2}{16}(49-9\sqrt{17}),\\
 c_{-1}=\frac{5\pi^2}{136}\bigl(17+7\sqrt{17}\bigr)\,c_1,\qquad c_{-2}=\frac{3\pi^4(1445+163\sqrt{17})}{1156}.
\end{gathered}
 \label{eq:coefficients}
\end{equation}
Numerically $c_1=3.990842\ldots$, $c_0=7.335613\ldots$, $c_{-1}=66.411910\ldots$, $c_{-2}=535.176892\ldots$. The constant correction was estimated by Richardson extrapolation in~\cite{Barletta2026}. The coefficients beyond $c_1$
follow from a routine continuation of the leading balance of~\cite{BarlettaCelli2017}, and the inversion formula of
section~\ref{sec:allorders} gives them at every order. The linear law has a simple meaning in terms of the packet edges.
At the threshold the flow speed equals the spreading speed, so $c_*(\Ra)\sim\Ra/c_1$: stronger heating lets the packet
spread against a faster flow. The constant and the inverse powers are diffusive corrections to this balance. The four terms give relative errors $0.1542\%$ at $\Pe=20$, $0.0101\%$ at $\Pe=50$ and $0.000697\%$ at $\Pe=100$.

\subsection{A general two-term law}\label{sec:resistance-asymptotic}

The same limiting saddle gives a compact formula for the whole class~\eqref{eq:general}. For fixed $A,B$,
\begin{equation}
 \mathcal R_{\rm a}=c_1\sqrt{AB}\,\mathcal P+\frac{\pi^2}{16}\bigl[(13-5\sqrt{17})A+4(9-\sqrt{17})B\bigr]+O(\mathcal P^{-1}).
 \label{eq:universal-asymptotic}
\end{equation}
It suffices to evaluate at the limiting saddle $\Rey\frac{\kappa_\infty^2}{1+\kappa_\infty^2}=\frac{9+\sqrt{17}}{16}$ and
$\Rey(1+\kappa_\infty^2)=\frac{9-\sqrt{17}}8$; a change of level by $s$ shifts the constant part of $r_s$ by $s/g_r=(9-\sqrt{17})s/4$, and
substitution of $s=B/A-1$ and the physical scales gives~\eqref{eq:universal-asymptotic}. The product $AB$ sets the leading
cost of holding the packet and the ratio $A/B$ its spatial scale, $k_\infty=\pi\sqrt{A/B}\,\kappa_\infty$. For anisotropy
$R_{\rm a}\sim c_1\sqrt a\,\Pe$, for a power-law fluid $R_{\rm a}\sim c_1\sqrt\nu\,\Pe^\nu$. In the Forchheimer model the
coefficients grow with the flow: for fixed $\mathit{Fo}>0$, $A\sim2\mathit{Fo}\,\Pe$, $B\sim\mathit{Fo}\,\Pe$, $s\to-1/2$, and
\begin{equation}
 \Ra_{\rm a}\sim\sqrt2\,c_1\mathit{Fo}\,\Pe^2 .
\end{equation}
This quadratic law and its coefficient $5.64390\ldots$ are given in equation~(8.75) of~\cite{Barletta2019}. The growth function shows how the linear, power and quadratic laws arise from one saddle.

\subsection{All orders of the large series}\label{sec:allorders}

On the elliptic curve the large series can be inverted once to give a formula of any order. Put
\begin{equation}
 X=\frac{c_1}{\pi},\qquad \tau=\frac{\pi^2}{c_1\Pe}=\frac1{Xp},\qquad \frac{\Ra_{\rm a}}{c_1\Pe}=\sum_{n=0}^{\infty}b_n\tau^n,\qquad
 c_{1-n}=\pi^{2n}c_1^{1-n}b_n .
 \label{eq:large-normalization}
\end{equation}
Near $u_\infty$ take $h_+(u)$ and the root $\sqrt{h(u)/h_\infty}$ equal to one at $u=u_\infty$, and define
\begin{equation}
 J_L(u)=\frac{11\sqrt{17}-3}{32}(1+2u)\sqrt{\frac{h(u)}{h_\infty}}\,[1+u-h(u)],\qquad
 G_L(u)=\frac{\mathcal B(u)}{4J_L(u)},\qquad H_L(u)=\frac{\mathcal A(u)^2}{J_L(u)} .
\end{equation}
Then $\tau=G_L(u)$ and the required function is $H_L(u(\tau))$, with $G_L(u_\infty)=0$, $H_L(u_\infty)=1$ and
$G_L'(u_\infty)=(187+3\sqrt{17})/128\ne0$. The Lagrange--B\"urmann formula~\cite{FlajoletSedgewick2009} gives, for $n\ge1$,
\begin{equation}
 b_n=\frac1n[\xi^{n-1}]\,H_L'(u_\infty+\xi)\left[\frac{\xi}{G_L(u_\infty+\xi)}\right]^n ,
\end{equation}
where $[\xi^{m}]$ denotes the coefficient of $\xi^m$. All coefficients lie in the ring
\begin{equation}
 b_n\in\mathbb Z[1/34,\sqrt{17}] .
\end{equation}
The reason is arithmetic. The elements inverted in the expansion about the limiting point have norms $N(u_\infty)=-1/4$
and $N(h_\infty)=-1/2$, so only powers of $2$ enter there. Formal inversion then determines each new coefficient by
division by $G_L'(u_\infty)$, whose norm is $17/8$. Hence $2$ and $17$ are the only primes in the denominators. The first 13 coefficients
obtained by inversion were checked independently by direct substitution into $F$.

\subsection{Two exact radii}

\begin{figure}[H]
\centering\includegraphics[width=\textwidth]{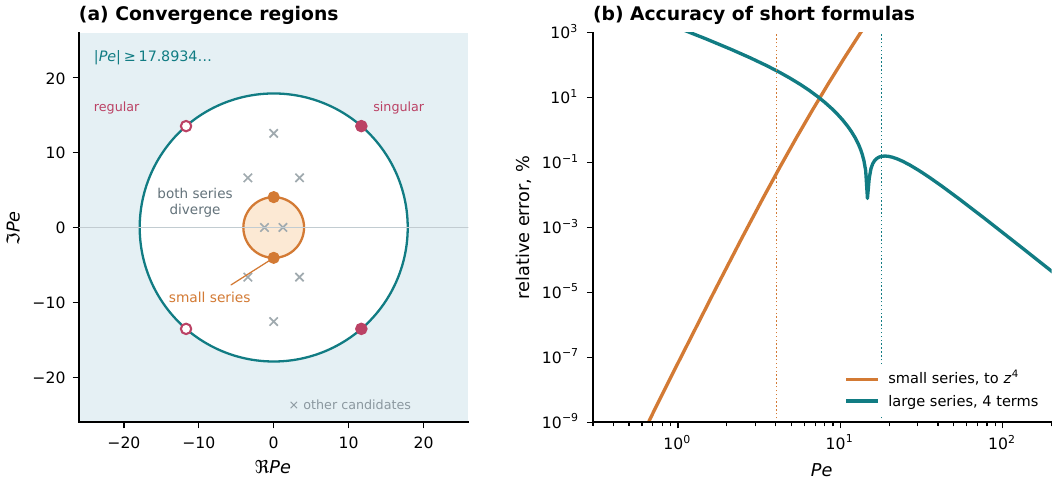}
\caption{Convergence and accuracy of the series. (a) The small series converges inside the small disk of the complex $\Pe$
plane and the large one outside the large disk. Filled points on the outer circle are singular for the selected large sheet,
open points are regular, crosses are other candidates from the discriminant. (b) Errors of short truncations relative to the exact threshold; the dip near $\Pe\approx14.8$ is a sign change of
the large-series error. Vertical lines mark the radii of convergence; the threshold remains analytic. Between them both series diverge and the
threshold is given by $F=0$, the parametrization~\eqref{eq:radical} or Algorithm~\ref{alg:onset}. The figure was drawn with the AI-assisted Python code described under \hyperref[sec:computational-methods]{Computational methods}.}
\label{fig:convergence}
\end{figure}

The small and the large series are limited by different complex singularities of the same curve
(figure~\ref{fig:convergence}(a)). Let
\begin{equation}
 \chi_\pm=-\frac{787\pm99\sqrt{33}}{578},\qquad \mu_\pm=\frac{85518\pm14894\sqrt{33}}{2601},\qquad
 P_s=\frac{\chi_--\sqrt{\chi_-^2-4\mu_-}}2<0 .
 \label{eq:radius-constants}
\end{equation}

\begin{theorem}[Radii on the selected sheets]\label{thm:radii}
The small series~\eqref{eq:small} has the exact radius $|z_s|$, set by its singularity at
$z_s=P_s/4<0$:
\begin{equation}
 |z_s|=-P_s/4=0.104007001009\ldots,\qquad \Pe_s=2\pi\sqrt{-P_s}=4.052669154822\ldots ,
 \label{eq:small-radius}
\end{equation}
and converges absolutely for $|\Pe|\le\Pe_s$. The large series~\eqref{eq:large-normalization}, continuing the positive branch
of the threshold, has radius $|\tau_c|=1/(2X\mu_+^{1/4})=0.138210845046\ldots$, set by its
complex-conjugate singularities $\tau_c$, that is
\begin{equation}
 \Pe_c=2\pi\mu_+^{1/4}=17.893405923407\ldots ,
 \label{eq:large-radius}
\end{equation}
and converges absolutely for $|\Pe|\ge\Pe_c$.
\end{theorem}

Figure~\ref{fig:convergence}(a) shows the regions of convergence. Figure~\ref{fig:convergence}(b) compares short
truncations with the exact threshold.

The complete proof, with the rational certificates, is given in appendix~\ref{app:certificates}; it has two steps. The discriminant lists the possible critical values; one must then establish which of them belong to the selected sheet.
This is the third selection step of the paper, after the saddles in section~\ref{sec:selection} and the frequency in
section~\ref{sec:algebraic}. At the critical value $0<P_a=0.038366588824\ldots<|P_s|$ other branches merge, while the selected branch
remains regular, so the radius cannot be read off from the zeros of the discriminant alone. For the small series the singularity
lies at negative $z_s=P_s/4$, so the coefficients eventually alternate and have order $|z_s|^{-n}n^{-5/2}$. The ratio of moduli of neighboring coefficients tends to $9.614737\ldots$. The large series has two complex-conjugate singularities
$\tau_c=0.090388\ldots\pm0.104557\ldots\ii$, which produce oscillations of the coefficients with the same power correction $n^{-5/2}$. The exponent comes from ordinary cusps of the algebraic projection, whose local expansion has a nonzero term of
power $3/2$~\cite{FlajoletOdlyzko1990}.

{\interlinepenalty=10000
At these cusps, $D=D_\kappa=D_{\kappa\kappa}=0$ and
$D_{\kappa\kappa\kappa}\ne0$: three spatial roots coalesce at an EP3. These are cusps of the projected curve $F=0$; they should not be
confused with the cusp points of the map $\kappa\mapsto\Lambda$ in the frequency plane~\cite{KupferBersRam1987}, which are
images of simple saddles. Such third-order spatial degeneracies were described
as super branch points in rotating-disk flow~\cite{Healey2004} and discussed by \'O~N\'araigh and
Spelt~\cite{ONaraighSpelt2013}. In the Prats problem their complex locations limit the selected threshold series; no such
coalescence occurs for real $r>1$, $p>0$.\par}

On the positive real axis the saddle stays simple and the threshold analytic at $\Pe_s$ and $\Pe_c$; both radii are
properties of the series alone. These numbers answer a computational question: how far the small series can be used, and from which throughflow the large
one converges.

\section{What carries over to general rational symbols, and limitations}\label{sec:general}

The argument has three model-dependent steps. (i) Locate all saddles and show that
they stay simple on the parameter domain. (ii) Compute the intersection indices at one convenient point (here $p\downarrow0$,
where the physical pair emerges from the real temporal maxima) and prove that they cannot change: no coalescence, and no
Stokes phenomenon involving a contributing saddle. (iii) Show that the selected growth is strictly monotone in the forcing
parameter, so the threshold is a unique root; elimination and geometry then give the exact boundary. For the symbols of
section~\ref{sec:universal} step (ii) rests on the reflection symmetry $L(-\overline\kappa)=\overline{L(\kappa)}$, which turns a
possible Stokes phenomenon of the off-axis pair into a perfect square of the spatial quartic, and on the phase structure of
the imaginary saddles. For a general rational symbol the first question is therefore the global structure of the physical saddles. Where a symmetry argument is unavailable, rational certificates of the kind used in appendix~\ref{app:certificates}
can certify the continuation of roots, while the control of Stokes alignments and of relative homology remains separate.

For the power-law fluid, the results apply to transverse rolls. The competition with longitudinal and
oblique rolls~\cite{AlvesBarletta2015} involves two horizontal wavenumbers and is not covered. In a rectangular anisotropic
channel the spanwise modes, $r\ne0$ in~\cite{CelliBarletta2022}, give a two-parameter family of symbols outside the class. Rheologies with a time-dependent momentum response change the structure of the symbol.

A horizontal temperature gradient also changes the symbol. With symmetric wall heat fluxes and an inclined
throughflow~\cite{Celli2025}, the dispersion relation comes from a numerical eigenvalue problem. Admissibility
has so far been checked there in sample cases. For three-dimensional localized disturbances in a layer with an inclined temperature gradient and
vertical throughflow, see Brevdo~\cite{Brevdo2009}. A certified global selection for numerically
defined dispersion relations would be the natural next step.
Finally, the analysis is linear and concerns unbounded layers. Relating the local absolute threshold to global modes or nonlinear fronts in finite
domains~\cite{HuerreMonkewitz1990,Chomaz2005,SandstedeScheel2000} requires separate treatment, as do porous media beyond the Darcy description~\cite{NieldBejan2017}.

\section{Conclusions}\label{sec:conclusions}

For Darcy flow, a horizontal throughflow leaves the onset of growth unchanged and moves the growth away from a fixed
observer. The temporal threshold remains $4\pi^2$, whereas the absolute threshold increases with the throughflow. At the
absolute threshold the trailing edge of the wavepacket stops, while the crests keep moving.

The same off-axis pair determines the large-time growth of the Green function throughout the domain $r>1$, $p>0$.
The exact threshold can then be computed in three ways: as the root of a monotone
function, as the selected branch of $F=0$, or along the elliptic arc. The polynomial gives exact crossings, the arc covers all throughflows, and the two series cover weak and strong throughflow. The certified radii show where each series stops converging, while the threshold itself stays analytic. The examples show three sources of ambiguity: noncontributing saddles, roots added by frequency
elimination, and neutral imaginary saddles below the threshold.

The porous convection models share one growth function. For transverse rolls, Forchheimer drag, the power-law index,
permeability anisotropy and magnetic resistance enter through the two coefficients of the linearized drag.
An open boundary changes the admissible vertical wavenumbers, and interphase heat exchange shifts the level in the
metallic foam model. At fixed throughflow the absolute threshold increases
with either drag coefficient. This explains why a vertical magnetic field raises the threshold at prescribed
$\Pe$, even when curves plotted against the imaginary saddle wavenumber cross.

Table~\ref{tab:formulas} collects the closed-form results of the paper.

\begin{table}[H]
\centering\small
\begin{tabularx}{\textwidth}{@{}lL@{}}
\toprule
Quantity & Closed form\\
\midrule
Temporal threshold & $\Ra_T=4\pi^2$~\cite{Prats1966}; $\mathcal R_T=\pi^2(\sqrt A+\sqrt B)^2$
by scaling, with the anisotropic and magnetic forms in~\cite{CelliBarletta2022,Turkyilmazoglu2023}\\
Neutral saddle locus & Equation and radical parametrization for Prats flow and a vertical magnetic field~\cite{Turkyilmazoglu2023};
here: quadratic form in $h$ for all levels and selection of the physical arc (on 11a3 at $s=0$)\\
Weak throughflow & Leading $1+z$, $z=\Pe^2/(16\pi^2)$~\cite{DufourNeel1998,ChungKimChoi2009};
here: $\Ra_{\rm a}/(4\pi^2)=1+z-z^2+3z^3-\cdots\in\mathbb Z[[z]]$\\
Strong throughflow & $c_1$ known~\cite{BarlettaCelli2017,Barletta2019,Turkyilmazoglu2023}, $c_0$ estimated in~\cite{Barletta2026};
exact $c_0,c_{-1},c_{-2}$ in~\eqref{eq:coefficients}, all $b_n$ by
Lagrange--B\"urmann, $b_n\in\mathbb Z[1/34,\sqrt{17}]$\\
Limiting saddle & $\kappa_\infty=\tfrac14\sqrt{7+\sqrt{17}}-\tfrac\ii4\sqrt{5+3\sqrt{17}}$~\cite{BarlettaCelli2017}\\
General two-term law & Equation~\eqref{eq:universal-asymptotic}; the leading Forchheimer and magnetic
terms are known from~\cite{Barletta2019,Turkyilmazoglu2023}, respectively\\
Crossing with $\Ra=\Pe^2/4$ & numerical value~\cite[pp.~192--193]{Barletta2019}; here: $\Pe_\times=\tfrac\pi2\sqrt{71+17\sqrt{17}}=18.6583\ldots$,
$\kappa_\times=-\ii\,\overline{\kappa_\infty}$\\
Projection node & new: $(\Pe,\Ra)=(63.78756623,\,263.08495654)$, crossed smoothly\\
Radical switch & new: $4u_D^3+4u_D^2-1=0$, $\Pe_D=34.216\ldots$\\
Radii & new: $\Pe_s=2\pi\sqrt{-P_s}=4.0527\ldots$, $\Pe_c=2\pi\mu_+^{1/4}=17.8934\ldots$ (theorem~\ref{thm:radii},
appendix~\ref{app:certificates})\\
Limits in the class & $r_s(0)=(1+\sqrt{1+s})^2$, the scaled temporal threshold~\cite{CelliBarletta2022,Turkyilmazoglu2023};
anisotropy $a\to0$: $R_{\rm a}\to\pi^2$~\cite{CelliBarletta2022}; open boundary
$\Ra^{\rm BC}_{\rm a}(\Pe)=\tfrac14\Ra_{\rm a}(2\Pe)$~\cite{BarlettaCelli2017}\\
\bottomrule
\end{tabularx}
\caption{Closed-form results and their relation to previous work.}
\label{tab:formulas}
\end{table}

\section*{Computational methods}\label{sec:computational-methods}
ChatGPT (OpenAI, GPT 6.1 Sol) and Claude (Anthropic, Claude Opus 5.5) assisted with algebraic derivations and
symbolic checks and with writing and debugging the numerical, verification, and plotting scripts. The author directed
this work by specifying the mathematical tasks and the checks to be performed, and reviewed the derivations, code,
and numerical results. The verification suite (\texttt{prats\_check.py}) tests exact algebraic identities,
compares series coefficients obtained by independent constructions, checks both convergence certificates using rational
arithmetic, and checks saddle selection by contour continuation and by direct quadrature of~\eqref{eq:green}.
It also compares the computed thresholds with published benchmarks.
The figures were drawn by executing the Python plotting scripts on the specified geometry, computed results, and
tabulated benchmarks.

\section*{Acknowledgments}
This work was supported by the British Academy (Pump Priming Collaboration between UK and EU Partners 2026, award
\mbox{PPHE26\textbackslash 100263}). The author acknowledges support of the Institut Henri Poincar\'e (UAR~839 CNRS-Sorbonne
Universit\'e), and LabEx CARMIN (\mbox{ANR-10-LABX-59-01}).

The author thanks Michael Berry for suggesting a direct numerical evaluation of the Fourier \mbox{integral~\eqref{eq:green}}.

The AI tools identified under \hyperref[sec:computational-methods]{Computational methods} also assisted with drafting
and editing the text. The author reviewed and edited their output and takes responsibility for the final manuscript.

\section*{Declaration of interests}
The author reports no conflict of interest.

\section*{Data availability statement}
The Python scripts, benchmark data, and rational certificate files will be deposited on Zenodo with a persistent
identifier (DOI) upon publication.

\appendix
\section{Boundary polynomial and frequency recovery}\label{app:polynomial}

The independent algebraic computation uses $P=p^2/4$ and the polynomial
\begin{equation}
\begin{split}
 F(r,P)={}&1024P^5-107P^4r^2-6464P^4r-4096P^4-8P^3r^4+740P^3r^3\\
 &+14564P^3r^2-40832P^3r+6144P^3+48P^2r^5-1824P^2r^4-7664P^2r^3\\
 &+63104P^2r^2-70976P^2r-4096P^2-96Pr^6+2272Pr^5-7328Pr^4\\
 &-12384Pr^3+45952Pr^2-12800Pr+1024P+64r^7-1024r^6\\
 &+5504r^5-10496r^4+3136r^3-256r^2 .
\end{split}
\label{eq:F}
\end{equation}
With $W=\Omega^2$,
\[
 \predisplaypenalty=10000
 \frac{\Delta(\ii\Omega)}{16}=\mathcal U(W)+\ii\Omega\,\mathcal V(W),\qquad
 \mathcal U(W)=e_2W^2+e_1W+e_0,\qquad \mathcal V(W)=W^2+f_1W+f_0,\postdisplaypenalty=10000
\]
with coefficients
\begin{equation}
\predisplaypenalty=10000
\begin{aligned} e_2&=P-4r+1,\\ e_1&=8P^2-3Pr^2+22Pr+8P+4r^3-22r^2+8r,\\ e_0&=16P^3+P^2r^2-40P^2r+16P^2-2Pr^3+32Pr^2-32Pr+r^4-8r^3+16r^2,\\ f_1&=3Pr+8P-6r^2+12r,\\ f_0&=-20P^2r+16P^2-Pr^3+32Pr^2-48Pr+r^4-12r^3+32r^2 .
\end{aligned}
\end{equation}
The common root is recovered as
\begin{equation}
 W=(r-P)\frac{K_W}{J_W},\qquad J_W=3P^2-15Pr-39P+20r^2-32r+4,
\end{equation}
\[
 K_W=-20P^2-Pr^2+91Pr-92P+4r^3-48r^2+132r-16 .
\]
When $J_W=0$ the common roots are found directly from the two quadratics; this covers the node and the crossing with $r=P$,
where the fraction may take the form $0/0$. The physical choice always requires $W>0$.

\section{Certification of the radii}\label{app:certificates}

The discriminant of $F$ with respect to $r$ factorizes as
\begin{equation}
 \operatorname{Disc}_rF=2^{36}P^3(P+4)^5Q_2(P)Q_4(P)^3N_6(P)^2 ,\postdisplaypenalty=10000
\end{equation}
where
\[
\begin{gathered}
 Q_2=8P^2+13P+16,\qquad Q_4=7803P^4+21249P^3+520020P^2+193536P-8192,\\
 N_6=9P^6+435P^5+18916P^4-16271616P^3-15442432P^2-22840320P+100352 .
\end{gathered}
\]
The factor $N_6^2$ corresponds to nodes of the projection, not to branching of the normalization; one of these nodes is
discussed in section~\ref{sec:algebraic}. For the choice of sheet it is convenient to eliminate $h$ from the elliptic model:
\begin{equation}
 K_7(u,P)=(4u^2+u-1)^2P^2+4u(2u-1)(2u+1)^2P+16u^3(2u+1)^4 .
 \label{eq:K7}
\end{equation}
The leading coefficient in $u$ is $256$, so no root escapes to infinity at finite $P$, and
\begin{equation}
 \operatorname{Disc}_uK_7=2^{36}P^9(P+4)(4P-9)^2Q_2Q_4 .
\end{equation}
The four relevant critical values come from the factorization
\begin{equation}
 Q_4=7803\,(P^2-\chi_+P+\mu_+)(P^2-\chi_-P+\mu_-).
\end{equation}
The pair with the plus sign is complex conjugate. The factor with the minus sign has two real roots $P_s<0<P_a$ with
$P_a=0.038366588824\ldots<|P_s|$; the closer point $P_a$ belongs to another sheet and does not limit the small series. At the
critical points of $Q_4$
\begin{equation}
 6u^2-3u-1=0 .
 \label{eq:cusp}
\end{equation}
On the smooth elliptic curve the first derivatives of $P$ and $r$ with respect to a local parameter vanish there, while the
vectors of second and third derivatives are linearly independent; at such a critical value $P_c$, after subtracting the analytic part, a nonzero term
$(P-P_c)^{3/2}$ remains.

To choose the sheet we use Rouch\'e's theorem~\cite{DLMFRouche}. Let $K(w_0+\xi)=\sum_j\beta_j\xi^j$. If on a disk of radius $\delta$
\begin{equation}
 |\beta_\ell|\delta^\ell>\sum_{j\ne\ell}|\beta_j|\delta^j ,
\end{equation}
the disk contains exactly $\ell$ roots counted with multiplicity. The coefficients are enclosed in rational intervals, and the
strict inequality is checked without floating-point rounding; neighboring segments are joined by a smaller disk containing the
same single root.

For the small series the substitution $u=Pv$ gives $K_7(Pv,P)/P^2=1-4v+O(P)$, so the physical germ has $v(0)=1/4$. A chain of rational segments reaches $P_a$ and keeps the simple positive root $u$; the roots merging there have $u=(3-\sqrt{33})/12<0$. Loops
around $P=0$ and $P=P_a$ do not change the selected root, and these are all the critical values inside $|P|<|P_s|$; its
monodromy is trivial and it is holomorphic in the whole disk. The function $r$ has no poles at finite $P$, because the leading
coefficient of $F$ in $r$ is constant. A further chain reaches the double root at $P=P_s$; the terminal disk contains
exactly two roots, and the check~\eqref{eq:cusp} establishes the nonzero term of power $3/2$. This proves that the small
radius is exact.

For the large series the substitution $P=1/(4X^2\tau^2)$ turns~\eqref{eq:K7} into a polynomial equation in $u,\tau$. The complex
roots of $Q_4$ have the largest modulus among the finite nonzero critical values and therefore give the smallest possible
nonzero modulus of $\tau$. For $0<|\tau|<|\tau_c|$ there are no multiple roots and the leading coefficient $256\tau^4$ does not vanish;
our germ is analytic at $\tau=0$, and a loop around the origin does not change it, so inside the disk the selected branch is
single-valued and holomorphic. A chain of rational segments brings it to one of the conjugate cusps, and a further chain proves regularity at the point with the opposite sign of $\tau$; the remaining two points follow by conjugation. The two certificates consist of $114$ and $403$ rational segments, with relative margin in Rouch\'e's inequality above
$1/500$. The script \texttt{prats\_certify.py} verifies the certificates. The exponent $3/2$ gives coefficients of order $n^{-5/2}$, so both series also converge absolutely on the
boundaries of their disks.

\begingroup
\small
\linespread{0.98}\selectfont

\endgroup
\end{document}